\documentclass[11pt]{article}
\usepackage{cite}
\usepackage{amsmath,amsfonts,amssymb}
\usepackage[small,bf,hang]{caption}
\usepackage{slashed}
\usepackage{color}
\usepackage{braket}

\usepackage{tikz-cd}

\usepackage{extarrows}
\usepackage{hyperref}

\def\hybrid{
        \topmargin -20pt
        \oddsidemargin 0pt
        \headheight 0pt \headsep 0pt
        \textwidth 6.25in 
        \textheight 9.5in 
        \marginparwidth .875in
        \parskip 5pt plus 1pt \jot = 1.5ex}

\hybrid

 \csname
@addtoreset\endcsname{equation}{section}

\def\moth{\mathsurround=0pt}
\newdimen\zo \zo=0pt

\def\tick{\leaders\hrule height 0.5ex depth 0pt \hskip 0.5pt}
\def\upboxfill{$\moth \setbox\zo\hbox{\tick}%
  \hskip 3pt\hbox to 0pt{$\tick$\hss}\hrulefill \hbox to 7.5pt{$\tick$\hss}$}

\def\dtick{\leaders\hrule height .34pt depth 0.5ex \hskip 0.5pt}
\def\downboxfill{$\moth \setbox\zo\hbox{\dtick}%
  \hskip 2pt\hbox to 0pt{$\dtick$\hss}\hrulefill \hbox to 2pt{$\dtick$\hss}$}

\def\bec{\begin{center}}
\def\ec{\end{center}}

\def\tr{{\rm tr}}

\def\be{\begin{equation}}
\def\ee{\end{equation}}
\def\bea{\begin{eqnarray}}
\def\eea{\end{eqnarray}}
\def\ba{\begin{array}}
\def\ea{\end{array}}

\begin{document}

\begin{titlepage}
\rightline{}
\rightline{July  2023}
\rightline{HU-EP-23/22}  
\begin{center}
\vskip 1.5cm
{\Large \bf{Holography as Homotopy}}
\vskip 1.7cm

{\large\bf {Christoph Chiaffrino, Talha Ersoy and Olaf Hohm}}
\vskip 1.6cm

{\it  Institute for Physics, Humboldt University Berlin,\\
 Zum Gro\ss en Windkanal 2, D-12489 Berlin, Germany}\\
\vskip .1cm

\vskip .2cm

ohohm@physik.hu-berlin.de, chiaffrc@hu-berlin.de,

\end{center}

\bigskip\bigskip
\begin{center} 
\textbf{Abstract}

\end{center} 
\begin{quote}  

We give an interpretation of holography in the form of the AdS/CFT correspondence in 
terms of homotopy algebras. 
A field theory  such as a bulk gravity theory can be viewed 
as a homotopy Lie or $L_{\infty}$ algebra. We extend this dictionary to theories defined on 
manifolds with a boundary, including the conformal boundary of AdS, 
taking into account the cyclic  structure needed to define an action with the correct boundary terms. 
Projecting fields to their boundary values then defines a homotopy retract, 
which in turn implies that the cyclic $L_{\infty}$ algebra of the bulk theory is equivalent, up to homotopy, 
to a cyclic $L_{\infty}$ algebra on the boundary. The resulting action is the `on-shell action' 
conventionally computed via Witten diagrams that, according to AdS/CFT, yields the generating 
functional for the correlation functions of the dual CFT. 
These results are established with the help of new techniques regarding the homotopy transfer 
of \textit{cyclic} $L_{\infty}$ algebras.

\end{quote} 
\vfill
\setcounter{footnote}{0}
\end{titlepage}

\tableofcontents


\newpage 

\section{Introduction}

Holography in the form of the AdS/CFT correspondence is arguably one of the most intriguing 
proposals that has emerged in quantum gravity. In its original and most tested form, the AdS/CFT conjecture 
relates the large $N$ limit of ${\cal N}=4$ super Yang-Mills theory with gauge group $SU(N)$ to 
type IIB supergravity on $AdS_5\times S^5$ \cite{Maldacena:1997re,Gubser:1998bc,Witten:1998qj}. 
Since then the AdS/CFT correspondence has been conjectured 
to extend to many other examples 
as well as to the full quantum regime, so that a theory of quantum gravity 
may be fully equivalent to a more or less conventional quantum field theory in flat space. 
While it  remains to be seen whether this strongest form of the  AdS/CFT conjecture 
really can provide a fully-fledged definition of quantum gravity on AdS, 
problems  of 
quantum gravity are now routinely phrased in the language of AdS/CFT.

Part of the psychological appeal of holography is undoubtedly due to the mystery of having two 
theories defined in different dimensions that are nevertheless  supposed to be fully  equivalent.
 We are of course familiar with  theories
that on the surface look quite different but that are secretly equivalent or dual to each other. 
Examples include  theories with a $p$-form that, for certain  matter couplings, 
are dual to theory with a $(D-p-2)$-form in $D$ dimensions. For instance, in four dimensions  a 
2-form may be dual to a scalar, so that in particular a theory with a gauge redundancy may be dual 
to a theory without gauge redundancy. For such examples the duality is easily established by writing a 
master action so that upon integrating out different fields one obtains either the original theory or its dual, 
thereby proving their equivalence. (See  \cite{Deser:1984kw} for an example of such proof.) 
However, in such familiar  examples the dual theories are defined in the same dimension, hence leaving 
it as somewhat of a mystery how holography could ever be proved from first principles.

In this paper we will provide a framework for AdS/CFT  in which the equivalence of  structures in different dimensions 
is built in. (See also the approach of Costello et.~al.~toward holography  \cite{Costello:2019jsy,Costello:2020jbh}.)  
This is the framework of homotopy algebras, which in pure mathematics arose in 
the context of algebraic topology \cite{stasheff} and in theoretical physics in string field theory 
\cite{Zwiebach:1992ie}. 
A classical perturbative field theory is encoded in a homotopy Lie algebra or $L_{\infty}$ 
algebra as follows \cite{Zwiebach:1992ie,Zeitlin:2007ttl,Zeitlin:2008cc,Zeitlin:2009zz,Hohm:2017pnh}: 
Denoting the fields collectively by $\phi$ one writes the action as 
 \be\label{IntroAction} 
  S[\phi] = \frac{1}{2} \omega(b_1(\phi), \phi) + \frac{1}{3!} \omega(b_2(\phi,\phi),\phi) + \frac{1}{4!} \omega(b_3(\phi,\phi,\phi),\phi) +\cdots \,,  
 \ee
with a potentially  infinite number of multi-linear maps  $b_n$ (with $n$ inputs), subject to generalized Jacobi identities 
defining an $L_{\infty}$ structure, see \cite{Hohm:2017pnh,Lada:1994mn,Lada:1992wc,Alexandrov:1995kv}.  
More precisely, the fields live in an integer graded vector space $V$ (in degree zero), 
which also contains gauge parameters (in degree $-1$), equations of motion or `anti-fields' (in degree $1$), and so on. 
Furthermore, in writing an action we actually assume a \textit{cyclic} $L_{\infty}$ algebra, with inner product  $\omega$ 
that  can be viewed as a symplectic form of odd degree. 
Such ingredients are familiar from the BV-BRST formalism \cite{Batalin:1981jr,Batalin:1984jr}, 
whose relation to cyclic $L_{\infty}$ algebras we will 
further clarify in this paper.

The usefulness of homotopy algebras is due to them behaving well under operations known 
as homotopy retract and homotopy transfer \cite{crainic2004perturbation,vallette2014algebra}. 
These are defined by projections from  the original graded vector space $V$ 
to a potentially much smaller space. Since such projections are generally not invertible, familiar algebraic structures such as Lie algebras 
are not preserved. In the larger category of homotopy Lie algebras, however, they are preserved provided the projection 
is invertible up to homotopy. Two homotopy algebras so related are equivalent in the context of homotopy Lie algebras, 
even if their underlying vector spaces have different dimensions. 
In particular, the cohomology, which  contains invariant physical information, is 
preserved, so that the resulting homotopy algebra on the smaller space may encode the same physical observables. 
It has been shown that homotopy transfer formalizes the physical notions of integrating out fields, see e.g.~\cite{Erbin:2020eyc,Koyama:2020qfb,Arvanitakis:2020rrk}, 
and passing over 
to gauge invariant variables \cite{Chiaffrino:2020akd} 
(in both cases projecting  from a larger 
to a smaller field space). Here we will show, inspired by earlier results in \cite{GwilliamThesis,GwilliamJF,Chiaffrino:2021pob,Cattaneo:2012qu}, 
that passing from a bulk theory on a manifold with boundary  to its boundary can be 
interpreted as a homotopy transfer. In a slight generalization of this result we show in particular  that passing from AdS to its conformal boundary 
also provides a homotopy transfer, hence proving that a theory on AdS, formulated as a homotopy algebra, 
may be equivalent to another homotopy algebra associated to the boundary.

In order to establish this result we have to refine  the  dictionary between field theory and cyclic $L_{\infty}$ algebras 
to take the boundary into account. Specifically, in order for (\ref{IntroAction}) to reproduce the correct action including boundary 
terms, the vector space $V$ has to be enlarged by `boundary components'. For instance, for a scalar field theory on 
a Riemannian manifold $M$
with boundary $\partial M$, the space of fields in degree zero is the space of smooth functions $C^\infty(M)$, 
but the space in degree one contains in addition  the space of smooth functions $C^\infty(\partial M)$ on the boundary. 
These spaces form the chain complex 
\begin{equation}
\begin{tikzcd}
0 \arrow[r] & V^0 = C^\infty(M) \arrow[r,"b_1"] & V^1= C^\infty(M) \oplus C^\infty(\partial M) \arrow[r] & 0 \, ,
\end{tikzcd}
\end{equation} 
where the only non-trivial differential $b_1$ is defined by 
\begin{equation}
b_1(\phi) = \left((-\Delta + m^2 ) \phi\,, \;\partial_n \phi|_{\partial M}\right) 
\, . 
\end{equation}
Here, $\Delta$ is the Laplace operator on $M$ (we are working in Euclidean signature), 
and $\partial_n$ is the derivative normal to the boundary.
Furthermore, we will show how to extend the complete cyclic $L_{\infty}$ algebra for scalar field theories and Yang-Mills theory, 
to this setting with boundary, 
which requires  adding an appropriate  boundary term to the symplectic form $\omega$. 

Starting then from  an $L_{\infty}$ algebra with boundary data included, we will define a homotopy retract 
by introducing the projection map $p$ from functions in the bulk to functions on the boundary. 
For a manifold $M$ carrying a boundary in the strict topological sense this projection is just the 
restriction to the boundary: $p(\phi)=\phi\big|_{\partial M}$.  
The prescription for AdS is slightly more subtle since here we have only a conformal boundary. 
Working in Poincar\'e coordinates where the metric for AdS$_{d+1}$ reads $ds^2=\frac{1}{z^2} (dz^2+dx^i dx^i)$, 
with $d$-dimensional  flat coordinates  $x^i$, the projection of a scalar $\phi(z,{x})$ 
includes a rescaling as follows: 
 \be
  p(\phi) := \lim_{z\rightarrow 0} z^{-\Delta_-} \phi\;, 
 \ee
where $\Delta_{-} = \frac{d}{2} - \sqrt{\frac{d^2}{4} + m^2}$. 
In both cases there is also an inclusion map $i$ in the other direction, i.e., 
from boundary functions to bulk functions. A given function $\phi_0({x})$ on the boundary is mapped by $i$ 
to the solution of the bulk equation $(-\Delta + m^2 ) \phi=0$ with boundary condition $\phi\big|_{\partial M}=\phi_0$ 
(or $\lim_{z\rightarrow 0} z^{-\Delta_-} \phi=\phi_0$ for AdS).  
The projection and inclusion maps are not inverse to each other since, for instance, the inclusion maps only to on-shell fields, 
while the space $V^0$ contains all off-shell fields. However, $i$ and $p$ are inverse `up to homotopy', 
which means that there is a degree $-1$ map $h$ so that $i\circ p = {\bf 1} - b_1\circ h - h\circ b_1$.  
This is sufficient to show that the $L_{\infty}$ algebra of the bulk  theory is transported to an equivalent $L_{\infty}$ algebra 
on the boundary.

Let us relate the above construction to the conventional AdS/CFT dictionary.  
Here one uses that according to the field/operator correspondence  
each operator ${\cal O}$ of the CFT corresponds to a field $\phi$ in the bulk theory whose  boundary value  $\phi_0$
 can be thought of as a `source' for ${\cal O}$ in that the generating functional or 
partition function is given by 
 \be\label{CFTpartitionfunction} 
  Z[\phi_0] = \Big\langle \exp\int d^4x\, \phi_0(x){\cal O}(x)\Big\rangle_{\rm CFT}\;. 
 \ee  
This function encodes arbitrary correlation functions involving ${\cal O}$, which can be obtained by taking 
suitable functional derivatives with respect to $\phi_0$. 
The AdS/CFT conjecture states that this partition function is computed 
by  the on-shell action of the bulk theory. Specifically, given the boundary value  $\phi_0$ 
one may find, using perturbation theory,  the solution $\phi(\phi_0)$ of the full non-linear bulk equations of motion that approaches 
$\phi_0$ at the boundary. The on-shell action is  obtained by inserting this solution 
into the bulk action, subject to a procedure  of `holographic renormalization' to render the result finite \cite{Skenderis:2002wp}. 
This on-shell action 
is conjectured to give the  partition function (\ref{CFTpartitionfunction}):
 \be
  Z[\phi_0] = \exp\big[-S[\phi(\phi_0)]\,\big]\;. 
 \ee
The on-shell action defines, using  (\ref{IntroAction}), a boundary cyclic $L_{\infty}$ algebra. 
It is this $L_{\infty}$ algebra that is computed by the above construction using homotopy transfer.  
Therefore, if AdS/CFT is correct, this $L_{\infty}$ algebra encodes all correlation functions of the CFT. 
This result follows from the  detailed construction of the homotopy transported $L_{\infty}$ brackets, 
which using the so-called perturbation lemma precisely implements the standard technique of Witten diagrams \cite{Witten:1998qj}, 
thereby giving a homotopy algebra interpretation of the latter.

The reader may well wonder what the point of the homotopy algebra formulation is, 
given that in the end one obtains the known method of Witten diagrams.  Apart from being of conceptual interest 
and giving, at least in our view, a more systematic introduction of these techniques, the perhaps most important 
point is that this formulation  provides a conceptual framework for holography 
(of structures  in different dimensions that are nevertheless equivalent), which hence 
may be a first baby step toward a first-principle derivation of AdS/CFT. (The methods presented here also appear to be related to 
the use of Hamilton-Jacobi theory for AdS/CFT \cite{deBoer:1999tgo,deBoer:2000cz}).

Let us then  try to imagine how a proof of the original AdS/CFT correspondence between 
${\cal N}=4$ super-Yang-Mills theory in four dimensions and type IIB supergravity on $AdS_5\times S^5$ could look like. 
We start with  the cyclic $L_{\infty}$ algebra of IIB supergravity on $AdS_5\times S^5$, which 
one could write down  quite explicitly, thanks to techniques that were developed during the last decade 
within a framework known as exceptional field theory \cite{Hohm:2013pua,Hohm:2013uia}.\footnote{These techniques allow one not only to 
embed the  truncation to five-dimensional gauged supergravity on $AdS_5$
 into the full ten-dimensional theory \cite{Hohm:2014qga,Baguet:2015sma}, 
but also to encode all massive Kaluza-Klein modes corresponding to the higher harmonics on $S^5$ \cite{Malek:2019eaz}, 
which are dual to operators on the CFT side and hence need to be included.} 
The results of this paper establish a homotopy transfer from this $L_{\infty}$ algebra to the 
$L_{\infty}$ algebra of the generating functional on the boundary $\mathbb{R}^4$, 
and hence an $L_{\infty}$ morphism  between them. This is indicated in the diagram as a solid line. 

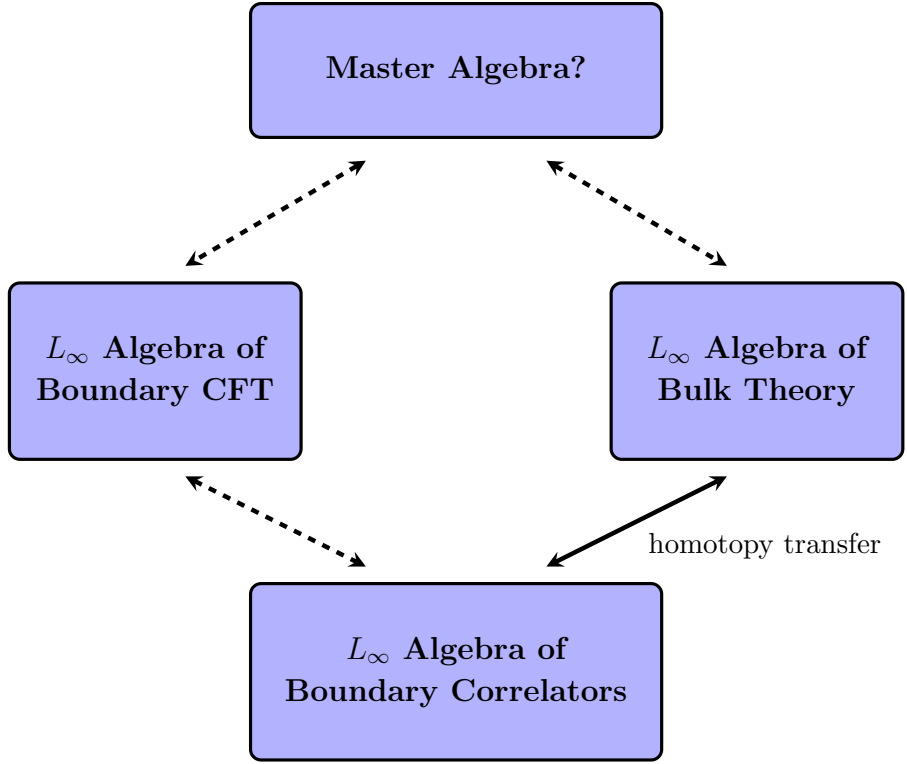
\tikzstyle{mybox} = [draw=black, fill=blue!30, very thick,
    rectangle, rounded corners, inner sep=10pt, inner ysep=20pt]
\begin{center}
\begin{tikzpicture}
\node [mybox] at (0,0) (box1){
    \begin{minipage}{0.3\textwidth}
	\begin{center} \large \textbf{Master Algebra?} \end{center}
    \end{minipage}
};
\node [mybox] at (-4,-4) (box2){
    \begin{minipage}{0.2\textwidth}
	\begin{center} \large \textbf{$L_{\infty}$ Algebra of Boundary CFT} \end{center} 
    \end{minipage}
};
\node [mybox] at (4,-4) (box3){
    \begin{minipage}{0.2\textwidth}
	\begin{center} \large \textbf{$L_{\infty}$ Algebra of Bulk Theory} \end{center}
    \end{minipage}
};
\node [mybox] at (0,-8) (box3){
    \begin{minipage}{0.3\textwidth}
	\begin{center} \large \textbf{$L_{\infty}$ Algebra of Boundary Correlators 
	} \end{center}
    \end{minipage}
};
\draw[stealth-stealth,dashed,line width=.6mm] (-1.2,-1.2) -- (-3.6,-2.6);
\draw[stealth-stealth,dashed,line width=.6mm] (1.2,-1.2) -- (3.6,-2.6);
\draw[stealth-stealth,line width=.6mm] (3.6,-5.4) -- (1.2,-6.6) node [midway, below right] {homotopy transfer};
\draw[stealth-stealth,dashed,line width=.6mm] (-3.6,-5.4) -- (-1.2,-6.6);
\end{tikzpicture}
\end{center}

In addition, one requires a map from the $L_{\infty}$ algebra of ${\cal N}=4$ super-Yang-Mills theory
to the generating functional of its correlation functions in the large $N$ limit. 
In this case we do not yet know of a homotopy algebra formulation of this map, hence being indicated by a dashed line 
in the diagram, but given 
results in \cite{Kajiura2003,Munster:2011ij,Doubek:2017naz,Arvanitakis:2019ald,Jurco:2019yfd,Okawa:2022sjf,Konosu:2023pal} concerning the closely related  computation of scattering amplitudes and quantum expectation values 
more generally \cite{Chiaffrino:2021pob}, 
it is plausible that there is an $L_{\infty}$ morphism in 
this direction. Supposing this to be the case, and using that $L_{\infty}$ morphisms are invertible 
up to homotopy, 
one would have an explicit $L_{\infty}$ morphism between 
${\cal N}=4$ super-Yang-Mills theory  and type IIB supergravity on $AdS_5\times S^5$ \textit{provided} 
the AdS/CFT conjecture is correct. In order to actually prove the AdS/CFT conjecture one would 
presumably have to define a master homotopy algebra, indicated by a question mark in the diagram, 
from which both the ${\cal N}=4$ super-Yang-Mills and the type IIB supergravity $L_{\infty}$ algebras 
can be obtained by homotopy transfer. Finding this master homotopy algebra, which could be 
quite abstract and may not be an  $L_{\infty}$ algebra itself, and hence proving AdS/CFT is of course a
tall order. But we take it as progress that one can now at least imagine how a proof could look like.

In the remainder of this introduction we outline the content of each section and emphasize our new 
technical results: 

In section 2 we review  the formulation of $L_{\infty}$ algebras based on the 
symmetric algebra $S^c V = \bigoplus_{n \ge 1} V^{\wedge n}$ of the underlying vector space $V$, 
on which one defines the coderivation $B=b_1+b_2+b_3+\cdots$ whose squaring to zero, $B^2=0$, encodes 
the generalized Jacobi identities. This formulation is well-known, but we here provide some important additions 
for the case of \textit{cyclic} $L_{\infty}$ algebras with symplectic form $\omega$ and their homotopy transfer. 
This allows us in particular to clarify the relation to the BV formalism, which is conventionally presented 
as an odd generalization of symplectic geometry, with a homological vector field or derivation squaring to zero 
that is dual to the coderivation. We show that the multilinear map given by $S=\omega(\pi_1B\otimes {\bf 1})$, 
with $\pi_1$ the projection operator from  $S^c V$  to $V$, gives 
the classical action (\ref{IntroAction}) when evaluated on fields in degree zero, but also gives  the complete 
BV master action when evaluated on the full vector space $V$. Finally, with these results, we prove that 
the homotopy transfer of the cyclic $L_{\infty}$ algebra yields an action that equals the `on-shell action' $\bar{S}$, 
i.e., $\bar{S} = S\circ i$, where $i$ is the perturbed inclusion map from initial or boundary conditions to 
full solutions of the `bulk' equations. We imagine these results to be useful beyond AdS/CFT.

In section 3 we discuss scalar field theories on Riemannian manifolds with boundary and 
establish homotopy transfer to the boundary. To this end we refine the  $L_{\infty}$  description for 
theories in presence of a boundary including the symplectic form, as outlined above, 
and show that the projector to the boundary, together with the trivial inclusion and a homotopy 
map given by the Green's function, satisfies the homotopy relation. 
We illustrate this for the one-dimensional harmonic oscillator, where the boundary is given 
by the endpoints of the finite time interval $[t_i, t_f]$.

In section 4 we generalize this construction to Yang-Mills theory, which we take to be a toy model for 
bulk supergravity. We introduce and review Hodge theory 
for manifolds with boundary and then specialize to Yang-Mills theory whose cyclic $L_{\infty}$  algebra 
is generalized to take the boundary into account, and 
we provide the homotopy transfer to the boundary.  In particular, 
using the multilinear map  $S=\omega(\pi_1B\otimes {\bf 1})$ we then automatically get the full BV action 
including boundary terms. 

Finally, in section 5 we apply, and slightly generalize, these results by considering scalar 
field theories on AdS. We establish the homotopy 
 to its conformal boundary, 
and compute the transported $L_{\infty}$ brackets on the boundary using the perturbation 
lemma. In this way we recover the method of Witten diagrams.

\section{Homotopy Transfer of Cyclic $L_\infty$ Algebras}

\subsection{Cyclic $L_\infty$ Algebras}\label{CyclicGeneralities}

\textit{Generalities:}\\[0.5ex] 
We give a short review of the definition of an $L_\infty$ algebra and how they arise in classical field theory, for more details see for example \cite{Lada:1994mn,Hohm:2017pnh}. An $L_\infty$ algebra on a graded vector space $V$ consists of a collection of graded symmetric maps
\begin{equation}
b_n: V^{\otimes n} \rightarrow V
\end{equation}
of degree one satisfying various relations. The first three of them are
\begin{enumerate}
\item $b_1^2 = 0$.
\item $b_1 b_2(x,y) = - b_2(b_1 x, y) - (-)^x b_2(x,b_1 y)$.
\item $b_1 b_3(x,y,z) + b_3 (b_1x,y,z) + (-)^x b_3(x,b_1 y, z) + (-)^{x + y}b_3(x,y,b_1 z) = \\ -b_2(b_2(x,y),z) - (-)^{(y+z)x}b_2 (b_2(y,z),x) - (-)^{(x+y)z}b_2(b_2(z,x),y)$.
\end{enumerate}
The first relation means that $b_1$ turns $V$ into a cochain complex, which is just a vector space with a square zero map of degree one. The second relation tells us that $b_1$ acts as a derivation on the product $b_2$. The last relation implies that $b_2$ fails to satisfy Jacobi, and this failure is measured by $b_1$ and $b_3$.

The infinite tower of relations among the $b_n$ can be encoded in a single equation in the following way. For this we need to introduce the symmetric graded coalgebra $S^c V$ over the graded vector space $V$. As a vector space, we have 
\begin{equation}
S^c V = \bigoplus_{n \ge 1} V^{\wedge n} \, ,
\end{equation}
i.e. it consists of polynomials over the elements of $V$ without constant term. In particular, the space $V^{\wedge n}$ consists of order $n$ monomials $x_1 \cdots x_n$, where $x_i \in V$, and we have graded symmetry among the $x_i$, i.e.
\begin{equation}
x_1 \cdots x_i x_{i+1} \cdots x_n = (-)^{x_i x_{i+1}} x_1 \cdots x_{i + 1}x_i \cdots x_n \, . 
\end{equation}
We then introduce a coproduct $\Delta: S^c V \rightarrow S^c V \otimes S^c V$. On order $n$ monomial it is defined as
\begin{equation}
\Delta(x_1 \cdots x_n) = \sum_{i = 1}^n \sum_{\sigma \in \bar{\text{Sh}}(i,n)} (-)^{\sigma(x_1,...,x_n)}x_{\sigma(1)}\cdots x_{\sigma(i)}\otimes x_{\sigma(i+1)}\cdots x_{\sigma(n)} \, .
\end{equation}
Here, the second sum runs over all $n$-element permutations $\sigma$, such that $\sigma(1) < ... < \sigma(i)$ and $\sigma(i+1) < ... < \sigma(n)$. The sign $\sigma(x_1,...,x_n)$ is then the natural sign one gets from reordering the $x_i$ with $\sigma$. The coproduct $\Delta$ is coassociative, i.e.
\begin{equation}
(\Delta \otimes 1) \Delta = (1 \otimes \Delta) \Delta.
\end{equation}
This is simply the dual relation to the associativity property of products $m: V\otimes V \rightarrow V$, which reads $m(m \otimes 1) = m(1\otimes m)$. We also need the concept dual to a derivation. We say that $B$ is a coderivation with respect to $\Delta$ if it satisfies
\begin{equation}
(B \otimes 1 + 1 \otimes B) \Delta = \Delta B \, .
\end{equation}

An $L_\infty$ algebra on $V$ can now be encoded in a degree one coderivation $B$ on $S^c V$, such that $B^2 = 0$. 
The data $(S^c V, \Delta, B)$ is an example of a differential graded coalgebra, which is the concept dual to that of a differential graded algebra. The $L_\infty$ products $\{b_n\}_{n \ge 1}$ are now given by
\begin{equation}\label{productsvscoder}
\sum_{n \ge 1} b_n := \pi_1 B \, ,
\end{equation}
where $\pi_1: S^c V \rightarrow V$ is the projection onto $V \subseteq S^c V$, i.e.~it extracts the linear part of a polynomial in $S^c V$. Also, $b_n$ in \eqref{productsvscoder} is the part of $\pi_1 B: S^c V \rightarrow V$ that acts on $V^{\wedge n}$. The infinite number of relations among the $b_n$ is now encoded in the single equation $B^2 = 0$. 

At this point we should mention that $B^2 = 0$ is equivalent to $\pi_1 B^2 = 0$, so the square zero condition does not encode further information over those imposed on the $b_n$. The equivalence holds because $B$ is a coderivation and not some general linear map. Coderivations on $(S^c V,\Delta)$ have the property that there is a one-to-one correspondence between linear maps $b: S^c V \rightarrow V$ and coderivations $B: S^c V \rightarrow S^c V$. In other words, for any such $b$ (which satisfies no additional property apart from being linear), there is a unique coderivation $B$, such that $\pi_1 B = b$. Also coderivations are closed under taking graded commutators, i.e.~for any two coderivations $A$ and $B$, $[A,B] := AB - (-)^{AB}BA$ is also a coderivation. For an odd $B$, we have $B^2 = \frac{1}{2}[B,B]$, so $B^2$ is also a coderivation independent of whether it is zero or not, and as such $B^2 = 0$ is equivalent to $\pi_1 B^2 = 0$ (the coderivation $0: S^c V \rightarrow S^c V$ obviously is a lift of the linear map $0: S^cV \rightarrow V$).

We can use $L_\infty$ algebras to encode the equations of motions and the gauge structure of a quantum field theory. A field $\phi$ in $V$ of degree zero satisfies the equations of motion if it is a solution to the Maurer-Cartan equation
\begin{equation}\label{MCequation} 
\sum_{n \ge 1} \frac{1}{n!}b_n(\phi,...,\phi) = 0 \, .
\end{equation}
The gauge transformations are given by
\begin{equation}
\delta \phi = \sum_{n \ge 0} \frac{1}{n!} b_{n+1}(\lambda,\phi,...,\phi) \, ,
\end{equation}
where $\lambda$ has degree $-1$ and serves as the gauge parameter. The $L_\infty$ relations then assure that the equations of motion are gauge covariant under these gauge transformations.

For $L_\infty$ algebras one defines two notions of equivalence between two different algebras. Suppose we have $L_\infty$ algebras $(V,\{b_n\}_{n \ge 1})$ and $(W,\{c_n\}_{n \ge 1})$. Let $B$ be a coderivation on $SV$ and $C$ a coderivation on $SW$ encoding the respective $L_\infty$ products. We say that $(V,\{b_n\}_{n \ge 1})$ is isomorphic to $(W,\{c_n\}_{n \ge 1})$, if there is a degree zero invertible coalgebra morphism $\Phi: (SV,\Delta_{SV}) \rightarrow (SW,\Delta_{SW})$, such that $\Phi B = C \Phi$. $\Phi$ being a coalgebra morphism means that
\begin{equation}
\Delta_{SW}\Phi = (\Phi \otimes \Phi) \Delta_{SV} \, .
\end{equation}
One can prove that $\Phi$ is invertible if and only if $\varphi = \pi_1 \circ \Phi|_{V}: V \rightarrow W$ is invertible. The second notion of equivalence is weaker. We say that a degree zero coalgebra morphism $\Phi: SV \rightarrow SW$ is a quasi-isomorphism, if $\varphi = \pi_1 \circ \Phi|_{V}: (V,b_1) \rightarrow (W,c_1)$ is a quasi-isomorphism. This means that $\varphi$ is a (co)chain map, i.e.~$\varphi b_1 = c_1 \varphi$, and $\varphi$ descends to an isomorphism on cohomologies. The most important result in the theory of $L_\infty$ algebras is that given a quasi-isomorphism $i: (W,\{c_n\}_{n \ge 1}) \rightarrow (V,\{b_n\}_{n \ge 1})$, there is always a quasi-isomorphism $p: (V,\{b_n\}_{n \ge 1}) \rightarrow (W,\{c_n\}_{n \ge 1})$, such that the linear pieces of $p$ and $i$ are inverse to each other on cohomology of $(V,b_1)$ and $(W,c_1)$. 

General coalgebra morphisms $S^c V \rightarrow S^c W$ satisfy a property we already encountered for coderivations. Namely, given any degree zero linear map $f: S^c V \rightarrow W$, there is a unique coalgebra morphism $F: S^c V \rightarrow S^c W$ such that $\pi_1 \circ F = f$. Explicitly, given $f: S^c V \rightarrow W$, the lift $F$ is given by
\begin{equation}
F = \sum_{n \ge 1} \frac{1}{n!}f^{\otimes n}\Delta^n \, ,
\end{equation}
where $\Delta^1 = 1$ and $\Delta^n = (1 \otimes \Delta^{n-1})\Delta$. The $n$th term in the sum is the part of $F$ that maps into $W^{\wedge n} \subseteq S^c W$.

The map $f: SV \rightarrow W$ can be viewed  as a non-linear  map $f: V \rightarrow W$ \cite{MarklNC}. 
 First of all in the ungraded case, given a point $x \in V$, we can associate the non-linear map 
\begin{equation}
\begin{split}
V \longrightarrow W \, ,   \qquad 
x \longmapsto \sum_{n \ge 1} \frac{1}{n!} f_n(x,\ldots ,x) \, ,
\end{split}
\end{equation}
with  $f_n = f\circ i_n$ and $i_n : V^{\wedge n} \hookrightarrow S^c V$ the inclusion. Here, the $f_n$ are the Taylor coefficients of the non-linear 
map $f$. One can further show that given $F: S^c U \rightarrow S^c V$ and $G: S^c V \rightarrow S^c W$ with corresponding non-linear maps $f: U \rightarrow V$ and $g: V \rightarrow W$, the non-linear map derived from $G \circ F$ is $g \circ f$. In other words, the association $F \mapsto f$ is an isomorphism of associative algebras, where the associative algebra structure in both cases is given by composition of maps.  This non-trivial fact is of course necessary for the interpretation of coalgebra morphisms in terms of 
non-linear maps to actually make sense.

In the graded case, it does not make sense to consider expressions like $f_n(x,\ldots ,x)$, since functions vanish on odd variables when $n \ge 2$. Nevertheless, in this case $f$ still defines a map of graded spaces $V \rightarrow W$. In the literature (especially in string field theory), one still finds expressions like $f_n(x,\ldots ,x)$ even in the graded case. This makes sense if there is a basis $e^i$ and a dual basis $e_i^*$, in which case one defines a ``super field'' $\Phi = \sum_i e^i e_i^*$. We can then write
\begin{equation}
f(\Phi) = \sum_{n \ge 1} \frac{1}{n!}f_n(\Phi,\ldots ,\Phi) \, .
\end{equation}
However, we actually have $\frac{1}{n!} f_n(\Phi,...,\Phi) = f_n$, hence $f(\Phi) = f$. The symmetry factor accounts for the fact that products of dual vectors act as
\begin{equation}
e_1^* \cdots e_n^* (a_1,\ldots ,a_n) = \sum_{\sigma \in S_n} \pm e_1^*(a_{\sigma(1)})\cdots e_n^*(a_{\sigma(n)}) \, ,
\end{equation}
where the sum runs over all permutations of $n$ elements and the sign is the natural one coming from the grading. Its symmetrization ensures that these products are well defined as functions on $V^{\wedge n}$. In this work we do not  want to assume the existence of a basis, and we therefore refrain from the use of super fields. 

From a physics perspective, coalgebra morphisms define maps between field spaces. Among these maps, quasi-isomorphisms preserve the relevant physical information. First of all, an isomorphism is just an invertible field redefinition. A quasi-isomorphism on the other hand also includes the possibility that some gauge degrees of freedom are fixed or that some fields are evaluated on-shell. Of course, also the inverse is possible, i.e.~that new gauge degrees of freedom are added or that auxiliary fields are introduced. \\

\noindent
\textit{Cyclic $L_\infty$ Algebras:}\\[0.5ex] 
A cyclic $L_\infty$ algebra is an $L_\infty$ algebra equipped with a non-degenerate inner product $\omega: V \otimes V \rightarrow \mathbb{C}$ of degree $-1$, which is anti-symmetric in its entries, i.e.
\begin{equation}
\omega(a,b) = -(-)^{(a-1)(b-1)}\omega(b,a) = -\omega(b,a) \, .
\end{equation}
The last equality follows by degree reasons: since the output of $\omega$ is of degree zero (it is a number), it is only non-zero for $|a|+|b|=1$. 
In addition, we require that the products $b_n$ are cyclic with respect to $\omega$, meaning that
\begin{equation}\label{Cyclicity}
\omega(a_0,b_n(a_1,...,a_{n})) = (-)^{a_n(a_0 + \cdots  + a_{n-1})}\omega(a_n,b_n(a_0,...,a_{n-1})) \, .
\end{equation}
Since the $b_n$ are graded symmetric, it follows that
\begin{equation}
\omega (b_n \otimes 1) : V^{\otimes (n + 1)} \longrightarrow \mathbb{C} 
\end{equation}
is graded symmetric in all entries.

Given an element $\phi \in V$ of degree zero, the inner product allows us to define the action
\begin{equation}
S[\phi] = \sum_{n \ge 1}\frac{1}{(n+1)!}\omega(b_n(\phi,...,\phi),\phi) \, .
\end{equation}
The variation of this action is given by
\begin{equation}
\delta S[\phi] = \sum_{n \ge 1}\frac{1}{n!}\omega(b_n(\phi,...,\phi),\delta \phi) \, ,
\end{equation}
where we used cyclicity. Since $\omega$ is non-degenerate, we deduce that the equations of motion are indeed the Maurer-Cartan equations (\ref{MCequation}).

The action $S$ as a function on $V^0 \subseteq V$ derives from the multilinear function $\omega ( \pi_1 B \otimes 1)$, which is well defined as a function on $S^c V$ by cyclicity of $B$, in the same fashion as general maps $V \rightarrow W$ derive from coalgebra morphisms in the ungraded case. Therefore, it makes sense to extend the action $S$ to a graded function on all of $V$ by defining
\begin{equation}
S: = \omega(\pi_1 B \otimes 1) \, .
\end{equation} 
Interestingly, this action defines a BV action, where the extension to non-zero degrees then also contains terms involving anti-fields and ghosts. 
(One can also derive an $L_\infty$ algebra from a BV action, which is perhaps 
more familiar, see e.g.~\cite{Grigoriev:2023lcc}). 
The anti-bracket can be defined via $\omega$. For this reason, $\omega$ is often called a linear symplectic form on $V$.

For cyclic $L_\infty$ algebras, we also need to refine the notion of a morphism. Given cyclic $L_\infty$ algebras $(V,\{b_n\}_{n \ge 1},\omega)$ and $(W,\{b_n'\}_{n \ge 1},\omega')$, a morphism between them is a map $f: SV \rightarrow V$, such that
\begin{equation}
\omega'(a,b) = \omega(f_1(a),f_1(b)) \ \, , \quad \omega(f_n(a_1,...,a_n),f_k(b_1,...,b_k)) = 0\ \, ,  
\end{equation}
for all $n + k > 2$. 

Below we need an alternative characterization of cyclicity. We claim that $B$ is cyclic if and only if there is a function $S: S^c V \rightarrow \mathbb{C}$, such that
\begin{equation} \label{AltCyc}
S \circ Y = \omega(\pi_1 B \otimes \pi_1 Y) \circ \Delta
\end{equation}
for all coderivations $Y$. We consider this form advantageous in proofs, since in this characterization we do not rely on an evaluation on elements $\phi \in V$.

To check the claim we first note that if $B$ is cyclic in the sense of \eqref{Cyclicity}, then
\begin{equation}
S = \omega(\pi_1 B \otimes 1)
\end{equation}
defines a function on $S^c V$. By the lifting property of coderivations, this in turn means that any coderivation $Y$ 
acts on $S$ as
\begin{equation}
Y = (1 \otimes \pi_1 Y) \circ \Delta \ .
\end{equation}
Therefore,
\begin{equation}
S \circ Y = \omega(\pi_1 B \otimes \pi_1 Y) \circ \Delta \ .
\end{equation}

The other direction is a little bit more involved. We first need the following. Let $F,G$ be maps on  $S^c V \otimes V \rightarrow \mathbb{R}$ (for example, $S = \omega(1 \otimes \pi_1 B)$ has this form independent of whether $B$ is cyclic or not). Then $F = G$ if and only if
\begin{equation}\label{1formeval}
F (1\otimes \pi_1 Y) \Delta = G (1 \otimes  \pi_1 Y) \Delta 
\end{equation}
for all coderivations $Y$. Obviously, if $F = G$, then the above equation holds. On the other hand, suppose \eqref{1formeval} holds. Without loss of generality, we can assume that $F$ and $G$ are of the form $F,G:  V^{\wedge n} \otimes V \rightarrow \mathbb{R}$. 
 Now we choose $Y$ to only have a linear piece $\lambda: V \rightarrow V$. Equation \eqref{1formeval} then says that
\begin{equation}\label{CyclicSum}
\sum_{\sigma \in \mathbb{Z}_n} \pm (F-G)(x_{\sigma(0)},...,x_{\sigma(n-1)}|\lambda(x_{\sigma(n)}))  = 0 \ , \quad x_i \in V \, , 
\; i = 0,...,n \ .
\end{equation}
The vertical line $|$ separates the $V$ input from the $V^{\wedge n}$ inputs. Let $E := \{e_i\}_{i \in I}$ 
be an arbitrary basis of $V$.\footnote{We use that every vector space, be it finite or infinite-dimensional, has a basis in the sense that 
every vector can be expressed as a finite linear combination of basis vectors. 
In this, and in life in general, we assume the axiom of choice.} 
We take $x_i \in E$ for all $i$. Let $k$ be the multiplicity of $x_n$ among the $x_i$, i.e. $k$ is the number of times in which we have $x_n = x_i$ for $i = 0,...,n$. Without loss of generality, we can assume that $k = 1$ when $x_0$ is odd, since otherwise at most one term in \eqref{CyclicSum} survives and equality holds for this choice of basis elements. We define $\lambda$ so that $\lambda(e_{j}) = e_{j}$, where $j \in I$ is such that $x_n = e_j$. We also define $\lambda(e_i) = 0$ for all $i \ne j$, so $\lambda$ extends to a unique function on all of $V$. With all these choices, equation \eqref{CyclicSum} reduces to
\begin{equation}
k(F-G)(x_0,...,x_{n-1}|x_n)  = 0 \ .
\end{equation}
From this it follows that $F = G$ on basis elements. Therefore, $F = G$ everywhere.

Coming back to the original goal to show that \eqref{AltCyc} implies cyclicity in the sense \eqref{Cyclicity},  assume that
\begin{equation}
S \circ Y = (-)^Y\omega(\pi_1 B \otimes \pi_1 Y) \circ \Delta \ .
\end{equation}
Now since $S$ always takes at least two  
elements, we can think of it as a function on $S^c V \otimes V$. In this case $Y$ acts as
\begin{equation}
S\circ Y = S(1 \otimes \pi_1 Y) \circ \Delta \ , 
\end{equation} 
where $S$ on the right hand side is thought of as a function of $V \otimes S^c V$. We just proved that this then implies that
\begin{equation}
S = \omega(\pi_1 B \otimes 1)
\end{equation}
as functions on $S^c V \otimes V$. But since $S$ is graded symmetric, so is $\omega(\pi_1 B \otimes 1)$. Therefore, $B$ is cyclic in the sense given in \eqref{Cyclicity}.

\subsection{Homotopy Transfer}

An important result in the theory of $L_\infty$ algebras is the homotopy transfer theorem. In physical theories, it allows us to gauge fix and to integrate out fields.

Suppose we have an $L_\infty$ algebra $(V,\{b_n\}_{n \ge 1})$ as well as a chain complex $(W,c_1)$ homotopy equivalent to $(V,b_1)$. This means that there are quasi-isomorphisms $p: (V,b_1) \rightarrow (W,c_1)$ and $i: (W,c_1) \rightarrow (V,b_1)$, as well as a degree $-1$ map $h: V \rightarrow V$ (the homotopy), such that
\begin{equation}
1 - ip = \{b_1, h\} \, .
\end{equation}
The homotopy transfer theorem then states that the chain complex structure on $(W,c_1)$ extends to an $L_\infty$ structure $(W,\{c_n\}_{n \ge 1})$, with the property that $i$ also has an extension to an $L_\infty$ morphism $i': (W,\{c_n\}_{n \ge 1}) \rightarrow (V,\{b_n\}_{n \ge 1})$, such that the linear piece of $i': S^c W \rightarrow V$ is $i$. In other words, $i'|_W = i$.

From a physics perspective, a homotopy equivalence of chain complexes $(V,b_1)$ and $(W,c_1)$ means that we have two equivalent \emph{free} theories. The transfer theorem then says then once we turn on interactions on $V$, we can find interactions for the theory on $W$ so that this interacting theory is still equivalent to the now interacting theory on $V$.  

Note that the homotopy transfer theorem does not guarantee us an extension for $p$. Although we mentioned above that a $L_\infty$ quasi-isomorphism always exists in the other direction, it is not necessarily an extension of $p$. In this work, we make some more assumptions, which then also give 
an extension $p'$ of $p$. These assumptions will be needed when we talk about homotopy transfer of cyclic algebras. 

To state the result using these stronger assumptions, we first recall the homological perturbation lemma and a version of it called the algebra perturbation lemma. The latter gives the result we need as a special case. 

The homological perturbation lemma assumes that we are given chain complexes $(X,\text d_X)$ and $(Y,\text d_Y)$, as well as quasi-isomorphisms $p: (X,\text d_X) \rightarrow (Y,\text d_Y)$ and $i: (Y,\text d_Y) \rightarrow (X,\text d_X)$ together with a homotopy $h: X \rightarrow X$, such that
\begin{equation}
1 - ip = \{\text d_X,h\} \, .
\end{equation}
Assume further that we have a perturbation $\delta: X \rightarrow X$ of the differential $\text d_X$, which means that we demand $(\text d + \delta)^2 = 0$. We further demand that $\delta$ is small in the sense that $(1 + \delta h)$ is invertible. Then we have  a differential $\text d_X'$ on $X$, quasi-isomorphisms $i': (X,\text d_X + \delta) \rightarrow (Y,\text d_Y')$ and $p': (Y,\text d_Y') \rightarrow (X,\text d_X + \delta)$, as well as a homotopy $h': X \rightarrow X$ satisfying
\begin{equation}
1 - i'p' = \{\text d_V + \delta, h'\} \, .
\end{equation}
Further, the perturbation lemma provides us with explicit formulae,
\begin{equation}
i' = 1 + h A i \, , \quad p' = p + p A h \, , \quad h' = h + h A h \, , \quad \text d_W' = \text d_W + p A i \, ,
\end{equation}
where $A = (1+\delta h)^{-1}\delta$. In our case, we assume that we can always write $A$ as a geometric series, i.e.~$A = \sum_{n \ge 1} (-\delta h)^n \delta$, which always works for our applications for  $L_\infty$ algebras, since in that case the sum will actually be finite when applied to an element in the symmetric tensor coalgebra.

In summary, the perturbation lemma states that a homotopy equivalence is stable under perturbations of the differential of $\text d_X$. To apply this to $L_\infty$ algebras, we set $X = S^c V$ and $Y = S^c W$ with differentials $\text d_X = B_1$ and $\text d_Y = C_1$, where $B_1$ and $C_1$ act on $S^c V$ and $S^c W$ as the coderivations so that they lift the linear differentials $b_1: V \rightarrow V$ and $c_1: W \rightarrow W$. We now assume that we have a homotopy equivalence $1 - ip = \{h,b_1\}$ between $V$ and $W$. From this we can construct a homotopy equivalence between $(S^c V,B_1)$ and $(S^c W,C_1)$ by lifting the maps $p$ and $i$ to cohomomorphisms $P$ and $I$, as well as defining
\begin{equation}
H(x_1 \cdots x_n) = \sum_{\sigma \in S_n}\sum_{i = 1}^n \pm \frac{1}{n!} x_{\sigma(1)}\cdots x_{\sigma(i-1)}h(x_{\sigma(i)})ip(x_{\sigma(i+1)})\cdots ip(x_{\sigma(n)}) \, .
\end{equation}
As always, the sign in front is the natural one coming from permuting the elements according to their degree. Also there are signs of the form $(-)^x$ coming from commuting elements $x$ past $h$. Having this set up, one can then show that
\begin{equation}
1 - IP = \{H,B_1\} \, .
\end{equation}
We can now apply the perturbation lemma using the perturbation $\delta = \sum_{n \ge 2} B_n$, where $B_n$ is the lift of the product $b_n: V^{\wedge n} \rightarrow V$. This then provides us with a differential $C'$ on $S^c W$ as well as with a homotopy eqivalence
\begin{equation}
1 - I' P' = \{h, B_1 + \delta\} \, .
\end{equation}

Unfortunately, the perturbation lemma does not guarantee that $C'$ is a coderivation on $S^c W$, so it is not guaranteed that $C'$ actually induces an $L_\infty$ structure. Further, it is not true in general that $I'$ and $P'$ are coalgebra morphisms. However, there is a variant of the homological perturbation lemma that guarantees these properties. For this, we further have to assume that
\begin{equation}\label{SideConditions}
H^2 = 0 \, , \quad HI = 0 \, , \quad PH = 0 \, , \quad PI = 1 \, ,
\end{equation}
which are sometimes called \emph{side conditions}. For us it is enough to assume these properties for $i$, $p$ and $h$, since then the above relations hold. If the side conditions  hold, we indeed have that $I'$ and $P'$ are coalgebra morphisms and $C'$ is a coderivation. In particular it follows that $C'$ is of the form $C' = \sum_{n \ge 1} c_n$, i.e.~we have $L_\infty$ products $c_n: W^{\wedge n} \rightarrow W$. Another important property is that the perturbed data still satisfies the side conditions.

For cyclic $L_\infty$ algebras we further require that
\begin{equation}
\omega(hx,y) = (-)^x \omega(x,hy)
\end{equation}
in addition to the side conditions. Note that this implies in particular that 
\begin{equation}\label{hionomega}
\omega (i \otimes h) = \omega(h \otimes i) = \omega(h \otimes h) = 0 \, .
\end{equation}
We first check that $I'$ is a cyclic $L_\infty$-morphism to $W$ with respect to the cyclic structure $\omega(i \otimes i)$. We have 
\begin{equation}
i' := \pi_1 I' = i -h \sum_{n \ge 1}(- \delta H)^n\delta I \, ,
\end{equation}
where on the right hand side $i$ and $h$ act as a zero on all $V^{\wedge n}$ for $n \ge 2$. Recall that $\delta = \sum_{n \ge 2} B_n$.  Using \eqref{hionomega}, it follows that
\begin{equation}
\bar \omega = \omega(i' \otimes i') = \omega(i \otimes i) \ .
\end{equation}
Therefore, $I'$ is indeed a cyclic $L_\infty$ morphism.

The last thing we have to prove is that the induced products $c_n$ are cyclic. For this we use our alternative characterization of cyclicity \eqref{AltCyc}. This means that we need to find a function $\bar{S}$, such that
\begin{equation}
\bar{S} = \omega(\pi_1 C' \otimes \pi_1 Y) \circ \Delta
\end{equation}
for all coderivations $Y$. We claim that $\bar{S} = S \circ I'$ has this property. We check this by computing $S\circ I' \circ Y$ for a given coderivation $Y$. For this, recall that $I'$ is given by
\begin{equation}
I' = \sum_{n \ge 0}\frac{1}{n!} {i'}^{\wedge n} \Delta^n \ .
\end{equation} 
Let $S_n$ the $n+1$-linear piece of $S$. We find that
\begin{equation}
\begin{split}
S_n\circ I' \circ Y &= \frac{1}{n!}S_n \circ {i'}^{\otimes n} \circ \Delta^{n} \circ Y = \frac{1}{(n-1)!} S_n \circ {i'}^{\otimes n}\circ (1^{\otimes {n-1}}\otimes Y)\circ \Delta^{n} \\
&= S_n \circ ( {i'}^{\otimes(n-1)}\otimes i' Y)(1 \otimes \Delta^{n-1})\circ \Delta = S_n \circ (I'\otimes i'Y) \circ \Delta \ .
\end{split}
\end{equation}
Here we used that $Y$ is a coderivation, which implies that
\begin{equation}
\Delta^n \circ Y' =\sum_{k+l = n-1} (1^{\otimes k} \otimes Y \otimes 1^{\otimes l})\circ \Delta^n \ , 
\end{equation} 
as well as the fact that $Y$ acts on a symmetric function, i.e.
\begin{equation}
F \circ \sum_{k+l = n-1} (1^{\otimes k} \otimes Y \otimes 1^{\otimes l}) =n F\circ  (Y \otimes 1^{\otimes (n-1)}) \ ,
\end{equation}
where $F$ is a function $F: (S^c V)^\otimes \rightarrow \mathbb{C}$ symmetric in each entry. 

Coming back to our derived expression for $S_n$, we now sum over all $n$, which then tells us that
\begin{equation}
S \circ I \circ Y = S \circ (I'\otimes i'Y) \circ \Delta \ . 
\end{equation}
We can now use that $S = \omega(\pi_1 B \otimes 1)$. We then conclude
\begin{equation}
\begin{split}
S \circ I \circ Y &= \omega(\pi_1 B I' \otimes i' \circ Y) \circ \Delta = \omega( \pi_1 I'C'  \otimes i' \circ Y) \circ \Delta \\ 
&= \omega(i \circ C' \otimes i \circ Y)\circ \Delta = (-)^Y\bar \omega(\pi_1 C'\otimes \pi_1 Y) \circ \Delta \ .
\end{split}
\end{equation}
Here we used that $B' I' = I' C'$ and $\omega(i' \otimes i') = \omega(i \otimes i) = \bar \omega$. This proves that $C'$ is cyclic.

Note that our proof also tells us that we do not  have to compute the $C'$ directly. Instead, we can compute $\bar \omega$ and $\bar S$ and find $C'$ from
\begin{equation}
\bar S = \omega(\pi_1 C' \otimes 1) \ .
\end{equation}

\section{Homotopy Retract to Boundary}
\subsection{Scalar Field Theory with Boundary}
\label{GenScalarField}

We first look at a  euclidean scalar field theory on some compact manifold $M$ with metric $g$ and with boundary $\partial M$. The action is given by
\begin{equation}\label{ScalarAction}
S = \int_M \sqrt{g}\left( \frac{1}{2} g^{\mu\nu}\partial_\mu \phi \partial_\nu \phi + \frac{1}{2} m^2 \phi^2 + V(\phi) \right)\, .
\end{equation} 
For the moment, we assume that there are no issues arising due to the presence of a boundary. In that case, the $L_\infty$ algebra has the underlying chain complex
\begin{equation}
\begin{tikzcd}
0 \arrow[r] & V^0 \arrow[r,"b_1"] & V^1 \arrow[r] & 0 \, ,
\end{tikzcd}
\end{equation}
where $V^0 = V^1 = C^\infty(M)$ and the differential is given by $b_1(\phi) = - \Delta \phi + m^2 \phi$, with $\Delta$ the Laplacian with respect to the metric $g$. The products $b_n$ depend on the form of the potential. For instance, for an $n$-point interaction $\frac{\lambda}{n!}\phi^n$ we have 
\begin{equation}
b_{k}(\phi_1,...,\phi_k)(x) =\begin{cases} \lambda\phi_1(x) \cdots \phi_{n-1}(x) & \text{for} \ k= n-1\\
0 & \text{otherwise} \, , 
\end{cases}
\end{equation}
so  that the Maurer-Cartan equation
\begin{equation}
\sum_{k \ge 1} \frac{1}{k!} b_k(\phi,...,\phi) = 0
\end{equation}
gives the correct equations of motion.

Let us denote by $\phi$ the fields in $V^0$ and by $\phi^* \in V^1$ the degree one objects, which we also call  \emph{anti-fields} due to their relation to the anti-fields in the BV formalism. If there is no boundary, the cyclic structure in that case is
\begin{equation}
\omega(\phi^*,\phi) := - \omega(\phi,\phi^*) := \int_M \sqrt{g} \phi^*\phi \, .
\end{equation}
This is only correct without a boundary, because  when one  computes the action via
\begin{equation}\label{CanonicalAction}
S' = \sum_{k \ge 1} \frac{1}{k!} \omega(b_k(\phi,...,\phi),\phi)\;, 
\end{equation}
one finds 
\begin{equation}\label{WrongScalarAction}
S' = \int_M \sqrt{g}\left(\frac{1}{2} g^{\mu\nu}\partial_\mu \phi \partial_\nu \phi + \frac{1}{2} m^2 \phi^2 + V(\phi)\right)  
- \frac{1}{2}\int_{\partial M} \sqrt{h} \phi \partial_n \phi \, , 
\end{equation}
where $\partial_n$ is the derivative in the direction normal to the boundary, and $h$ is the induced metric on the boundary. For simplicity we assume that the potential does not contain derivatives. Otherwise, there would also be boundary terms coming from the potential.

Due to boundary terms, the action $S'$ in \eqref{WrongScalarAction} is not equal to the action $S$ in \eqref{ScalarAction}. 
Furthermore,  
$b_1$ is not cyclic with respect to $\omega$. We could drop the requirement that we want a cyclic $L_\infty$ algebra and just consider a general  $L_\infty$ algebra, in which case all the physical information is encoded in the $b_k$, which is sufficient for  the equations of motion, 
but not the action. In order to encode an action we have  to modify the $L_\infty$ structure and/or the cyclic structure so that we obtain \eqref{ScalarAction} from \eqref{CanonicalAction}, which we do in the following. 

To this end we consider the general  variation of  the action \eqref{ScalarAction} but  keeping boundary terms: 
\begin{equation}
\delta S = \int \sqrt{g}\left[\delta \phi (- \Delta + m^2)  \phi + \delta \phi V'(\phi)\right]  + \int_{\partial M} \sqrt{h} \delta \phi \partial_n \phi \, .
\end{equation}
We interpret this as implying that apart from the `bulk' equations of motion
\begin{equation}
(- \Delta + m^2)  \phi +  V'(\phi) = 0\,, 
\end{equation}
there is also the `boundary equation of motion' 
\begin{equation}\label{boundaryeom}
\partial_n \phi\big|_{\partial M} = 0 \, . 
\end{equation}
This suggests to enlarge the space of antifields to $V^1 = C^\infty(M) \oplus C^{\infty}(\partial M)$ and 
 then to define
\begin{equation}
b_1'(\phi) = \left(-\Delta \phi + m^2 \phi\,, \; \partial_n \phi|_{\partial M}\right)\; .
\end{equation}
Next we should  account for the enlarged space in the definition of $\omega$. Denoting  by $\pi^*$ the fields in $C^{\infty}(\partial M)$, we define 
\begin{equation}
\omega'((\phi^*,\pi^*),\phi) := -\omega(\phi, (\phi^*,\pi^*)) := \int_M \sqrt{g} \phi^* \phi + \int_{\partial M} \sqrt{h} \pi^* \phi \, .
\end{equation}
With these choices it is now straightforward to show that $S$ in \eqref{ScalarAction} is given by
\begin{equation}
S = \sum_{n \ge 1} \frac{1}{(n+1)!} \omega'(b_n'(\phi),\phi) \, ,
\end{equation}
where the higher products $b_n$, $n\geq 2$,  are extended from $C^{\infty}(M)$ to $C^{\infty}(M) \oplus C^{\infty}(\partial M)$ trivially, i.e. $b_n'(\phi_1,...,\phi_n) = (b_n(\phi_1,...,\phi_n),0)$.\footnote{This can be extended to the case when $V$ does have derivative interactions: 
To this end note that for a general  Lagrangian density $\sqrt{g}\mathcal{L}(\phi,\partial \phi,x)$, 
the boundary term coming from the variation is always given by
\begin{equation}
\int_{\partial M} \sqrt{h} \frac{\delta \mathcal{L}}{\delta \partial_n \phi}\delta \phi \, .
\end{equation}
The `boundary equations of motion'  then tell us that the canonical momentum $\pi = \frac{\delta \mathcal{L}}{\delta \partial_n \phi}$ should vanish at the boundary. If the canonical momentum depends nonlinearly on $\phi$, we can account for this by modifying the higher products $b_n$. 
We will encounter this phenomenon later when we consider Yang-Mills theory.}

\subsection{Scalar Green's Function}

In this section we work out the properties of the scalar Green's function, which will define the homotopy of the bulk theory to the boundary theory. To an anti-field $\phi^*$, the Green's function associates a field $h(\phi^*)$, such that
\begin{equation}
(-\Delta + m^2)h(\phi^*) = \phi^* \, .
\end{equation}
Lemma 3.4.7 in \cite{Schwarz1995HodgeD} tells us that  $G$ is unique, once we choose Dirichlet boundary conditions 
$h(\phi^*)\big|_{\partial M} = 0$.

We will now  prove that $h$ defines a homotopy to the space of fields living on the boundary. In addition, since $h$ should give rise to a cyclic $L_\infty$ morphism, we need to prove the side conditions as well as symmetry with respect to $\omega$. To simplify notation, we define $L = -\Delta + m^2$ (we will not call it $b_1$, since $b_1$ now contains also the boundary equations of motion). We will frequently use Green's identity, which in our case is the failure of $L$ being cyclic with respect to $\omega$:
\begin{equation}
\omega(L \phi_1, \phi_2) + \omega(\phi_1 , L \phi_2) = \int_{\partial M} \sqrt{h}\Big\{\phi_1\partial_n \phi_2 - \phi_2 \partial_n \phi_1\Big\} \, . 
\end{equation}
This relation can be used to define a symplectic form  on phase space. To this end we define $\pi(\phi) = (\partial_n\phi|_{\partial M},\phi|_{\partial M})$, 
the projection of a field $\phi$ to its value and canonical momentum on the boundary. Further defining  the Hamiltonian symplectic form
\begin{equation}
\omega_{\partial} ((\varphi_1,\varpi_1),(\varphi_2,\varpi_2)) = \int_{\partial M}\sqrt{h}(\varphi_1 \varpi_2 - \varphi_2 \varpi_1)\,, 
\end{equation} 
we can write Green's identity as
\begin{equation}
\omega(L \phi_1, \phi_2) + \omega(\phi_1 , L \phi_2) = \omega_{\partial}(\pi(\phi_1),\pi(\phi_2)) \, .
\end{equation}
A non-linear version of this identity appears in the BV-BFV formalism of Cattaneo et.~al.~\cite{Cattaneo:2012qu}, where it characterizes the failure of the BV symplectic form to be invariant under the cohomological vector field. 

We begin by proving symmetry of $h$ with respect to $\omega$, computing 
\begin{equation}
\begin{split}
\omega(h\phi_1^*,\phi_2^*) + \omega(\phi_1^*,h\phi_2^*)  &= \omega(h\phi_1^*,Lh\phi_2^*) + \omega(Lh\phi_1^*,h\phi_2^*) \\ &= \omega_\partial(\pi (h\phi_1^*),\pi (h \phi_2^*)) \, .
\end{split}
\end{equation}
The right hand side vanishes if and only if $\omega_\partial$ vanishes on the image of $\pi h$. In the context of symplectic geometry, a subspace where the symplectic form vanishes is called isotropic. We can therefore say that $h$ is symmetric if and only if $h$ maps to an isotropic subspace on the boundary. In our case we demand that $h$ satisfies Dirichlet boundary conditions, so the fields in the image of $h$ will vanish on the boundary, but their canonical momentum can take arbitrary values. Fields satisfying the constraint $\phi|_{\partial M} = 0$ 
obviously specify an isotropic subspace. Hence, a propagator with Dirichlet boundary conditions is necessarily symmetric.

We next want to show that  $h$  generates solutions to the free bulk equations of motion. To this end  
we define a map $\iota: C^{\infty}(\partial M) \oplus C^{\infty}(\partial M) \rightarrow C^\infty(M)$ to be the adjoint of $\pi h$, in the sense
\begin{equation}\label{DefinitionOfInclusion}
\omega(\phi^*, \iota(\varphi,\varpi)) := \omega_{\partial}(\pi h \phi^*,(\varphi,\varpi)) \, .
\end{equation}
We show that the image of $\iota$ lies in the space of  solutions to $L\phi = 0$. Suppose we have a general field $\phi$.  We then find that
\begin{equation}\label{HomotopyRelationL1}
\begin{split}
\omega(\phi^*, \iota \pi(\phi)) &= \omega_\partial(\pi h(\phi^*),\pi \phi) = \omega(L h(\phi^*),\phi) + \omega(h(\phi^*), L \phi) \\ 
&= \omega(\phi^*,\phi) - \omega(\phi^*,hL(\phi)) \, ,
\end{split}
\end{equation}
where we used that $Lh = 1$ and that $h$ is symmetric with respect to $\omega$. Rewriting \eqref{HomotopyRelationL1}, we obtain 
\begin{equation}\label{HomotopyRelationL2}
\omega(\phi^*,hL(\phi)) = \omega(\phi^*,(1 - \iota \pi)(\phi)) \, .
\end{equation}
Since this has to hold for all $\phi^*$, we can deduce that 
\begin{equation}\label{HomotopyRelationL3}
hL = 1 - \iota \pi
\end{equation}
on fields $\phi \in C^{\infty}(M)$. Acting on both sides with $L$, we find that $L\iota \pi = 0$. But from this it follows that $L\iota = 0$, since $\pi$ is obviously surjective. Therefore, the image of $\iota$ are solutions to $L\phi = 0$, which we wanted to show. In fact, all solutions are in the image of $\iota$, since when $\phi$ is a solution, we see from \eqref{HomotopyRelationL2} that 
\begin{equation}\label{PisProjector}
\phi = \iota \pi (\phi) \, .
\end{equation}.

Let us come back to \eqref{DefinitionOfInclusion}. We can actually make an improvement there. Since $h$ satisfies Dirichlet boundary conditions, $\iota(\varphi,\varpi)$ will be independent of $\varpi$. We can therefore define a new map $i: C^{\infty}(\partial M) \rightarrow C^{\infty}(M)$ to be the restriction of $\iota$ to position variables. Likewise, we define a projection
\begin{equation}
\begin{split}
p: C^{\infty}(M) \rightarrow C^{\infty}(\partial M) \, , \qquad 
\phi \mapsto \phi|_{\partial M} \, .
\end{split}
\end{equation} 
The relation \eqref{HomotopyRelationL3} becomes
\begin{equation}\label{HomotopyRelationL4}
hL = 1 - i p \, .
\end{equation}
We prefer $i$ over $\iota$ because  $i$ is injective. This can easily be deduced from \eqref{HomotopyRelationL4} by acting with $p$ from the right and using $ph = 0$ (i.e.~$h$ satisfies Dirichlet boundary conditions). It then follows that $p = pip$, from which we can deduce that $pi = 1$, since $p$ is surjective. From $pi = 1$ we then deduce that $i$ is injective. Furthermore, for any boundary value $\varphi \in C^{\infty}(\partial M)$, we have that $pi(\varphi) = \varphi$, which means that $i$ associates to $\varphi$ the solution to $L\phi = 0$ with boundary value $\phi|_{\partial M} = \varphi$. In other words, $i(\varphi)$ solves the Dirichlet problem
\begin{equation}
L\phi = 0 \ , \qquad \phi|_{\partial M} = \varphi \, .
\end{equation}
In the context of AdS/CFT, $i$ is called the \emph{boundary-to-bulk} propagator, while $h$ is the \emph{bulk-to-bulk} propagator.

We can extend $p$ to anti-fields via $p(\phi^*) = 0$ and $h$ to fields via $h(\phi) = 0$. Also, we define $i$ to have zero image in anti-fields. The identity \eqref{HomotopyRelationL4} then reads
\begin{equation}\label{BulkHomotopy}
\{L,h\} = 1 - ip \, .
\end{equation} 
This means that we have a homotopy from the full complex defined by the differential $L: C^{\infty}(M) \rightarrow C^{\infty}(M)$ to the cohomology of $L$, which is represented by the boundary fields in $C^{\infty}(\partial M)$. Diagrammatically, this reads
\begin{equation}
\begin{tikzcd}
0 \arrow[r] & C^{\infty}(M) \arrow[r,"L"] \arrow[d,"p"] & C^{\infty}(M) \arrow[r] \arrow[d] & 0 \ , \\
0 \arrow[r] & C^{\infty}(\partial M) \arrow[r] & 0 \arrow[r] & 0 \ .
\end{tikzcd}  
\end{equation}
In fact, $h$ is a strong deformation retract (it satisfies the side conditions \eqref{SideConditions}). We trivially have $hi = 0$ and $h^2 = 0$. Further, we already saw that $pi = 1$ and $ph = 0$.

Summarizing, we have found a homotopy to solutions with respect to the differential $L: C^\infty(M) \rightarrow C^{\infty}(M)$.  In the next subsection we turn to the problem of extending  this homotopy to the differential $b_1: C^{\infty}(M) \rightarrow C^{\infty}(M)\oplus C^{\infty}(\partial M)$. 

\subsection{Homotopy Retract}

In the previous subsection we constructed maps $p: C^\infty(M) \rightarrow C^\infty(\partial M)$ and $i: C^\infty(\partial M) \rightarrow C^\infty(M)$ from fields defined on $M$ (bulk) to fields defined on $\partial M$ (boundary), together with a homotopy $h$ from $1$ to $ip$. We now want to extend this to the cyclic theory.

Recall from section \ref{GenScalarField} that in order to have a cyclic differential in the presence of a boundary, we extended the standard scalar field theory complex 
\begin{equation}
\begin{tikzcd}
0 \arrow[r] & C^{\infty}(M) \arrow[r,"L"] & C^{\infty}(M) \arrow[r] & 0
\end{tikzcd}
\end{equation}
to the complex
\begin{equation}
\begin{tikzcd}
0 \arrow[r] & C^\infty(M) \arrow[r,"b_1"] & C^\infty(M) \oplus C^\infty(\partial M) \arrow[r] & 0 \, ,
\end{tikzcd}
\end{equation} 
where 
\begin{equation}
b_1(\phi) = \left((-\Delta + m^2 ) \phi\,, \;\partial_n \phi|_{\partial M}\right) = \left(L(\phi)\,, \;p(\partial_n \phi) \right) \, .
\end{equation}
Here, $p$ is the projection to the boundary introduced in the previous section, and $\partial_n$ is the derivative normal to the boundary.

We would like to use the above  homotopy 
with respect to the differential $L$ 
also in the extended case, where the differential is given by $b_1$. In other words, we want a homotopy that solves the bulk equations of motion $L\phi = 0$, but leaves the boundary equations of motion $p(\partial_n \phi) = 0$. The complex in this case then looks like
\begin{equation}\label{ScalarBoundaryComplex}
\begin{tikzcd}
0 \arrow[r] & C^\infty(\partial M) \arrow[r,"c_1"] & C^\infty(\partial M) \arrow[r] & 0 \, ,
\end{tikzcd}
\end{equation}
where the differential $c_1$ computes $p_0(\partial_n \phi)$ on solutions to the bulk equation of motion. In the previous subsection, we defined the map $i: C^{\infty}(\partial M) \rightarrow C^{\infty}(M)$ that associates a solution to given boundary data. In terms of this boundary data, the differential $c_1$ is given by
\begin{equation}
c_1(\varphi) = p \partial_n i(\varphi) \, .
\end{equation}
In fact, due to the Hamilton-Jacobi equations this can also be written as $c_1(\varphi) = \frac{\delta (S \circ i)}{\delta \varphi}$. If we manage to find a cyclic quasi-isomorphism to the complex \eqref{ScalarBoundaryComplex}, this is of course what we would expect.

The first step to give a homotopy is to define a chain map between the bulk and boundary theory, which is a commuting diagram of the form
\begin{equation}\label{BoundaryCMap}
\begin{tikzcd}
0 \arrow[r] & C^\infty(M) \arrow[r,"\partial"] \arrow[d,"p_0"] & C^\infty(M) \oplus C^\infty(\partial M) \arrow[d,"p_1"] \arrow[r] & 0 \\
0 \arrow[r] & C^\infty(\partial M) \arrow[r,"\bar \partial"] & C^\infty(\partial M) \arrow[r] & 0 \, .
\end{tikzcd} 
\end{equation}
In degree zero, we use the restriction map $p_0 := p$ we are already familiar with. In degree one, we define
\begin{equation}
p_1(\phi^*,\pi^*) = \pi^* - p_0\circ \partial_n h(\phi^*) \, ,
\end{equation}
where $h$ is the Dirichlet Green's function. For $p_0$ and $p_1$ to combine into a chain map, we need $c_1 \circ p_0 = p_1 \circ b_1$, which we now show:
\begin{equation}
\begin{split}
p_1 b_1(\phi) &= p_1(L(\phi),p_0\partial_n \phi) = p_0\partial_n \phi - p_0\partial_n h L(\phi) \\
&= p_0\partial_n \phi - p_0(\partial_n - \partial_n i_0p_0)(\phi) =  c_1 p_0(\phi) \, .
\end{split}
\end{equation}
Therefore, $p_0$ and $p_1$ combine into a chain map
\begin{equation}
\begin{tikzcd}
0 \arrow[r] & C^{\infty}(M) \arrow[r,"b_1"] \arrow[d,"p_0"] & C^{\infty}(M) \oplus C^{\infty}(\partial M) \arrow[r] \arrow[d,"p_1"] & 0 \ , \\
0 \arrow[r] & C^{\infty}(\partial M) \arrow[r,"c_1"] & C^{\infty}(\partial M) \arrow[r] & 0 \ .
\end{tikzcd}
\end{equation}

For a fully fledged homotopy, we also need a chain map in the other direction. In degree zero, we take $i_0: C^\infty(\partial M) \rightarrow C^\infty(M)$. In degree one, we choose $i_1: C^\infty(\partial M) \rightarrow C^\infty(M) \oplus C^\infty(\partial M)$ to be the inclusion into the second factor. 
We prove  that $i_0$ and $i_1$ combine into a chain map:
\begin{equation}
b_1 i_0(\varphi) = (0,p_0\partial_n i_0(\varphi)) = i_1 c_1(\varphi) \, .
\end{equation}
Here, we used that $Li_0 = 0$. We therefore have a chain map
\begin{equation}
\begin{tikzcd}
0 \arrow[r] & C^{\infty}(M) \arrow[r,"b_1"] & C^{\infty}(M) \oplus C^{\infty}(\partial M) \arrow[r] & 0 \ , \\
0 \arrow[r] & C^{\infty}(\partial M) \arrow[u,"i_0"] \arrow[r,"c_1"] & C^{\infty}(\partial M) \arrow[r] \arrow[u,"i_1"] & 0  \ .
\end{tikzcd}
\end{equation}
Note also that 
\begin{equation}\label{PIis1}
p_0 i_0 = 1 \ , \qquad p_1 i_1 = 1 \ .
\end{equation}

The last step is to give a homotopy from $i \circ p$ to the identity. This is a map $h_1: C^\infty(M)\oplus C^\infty(\partial M) \rightarrow C^\infty(M)$ 
satisfying 
\begin{equation}\label{WhatGDoes}
h_1 b_1 = 1 - i_0 p_0\,, 
\end{equation}
and
\begin{equation}\label{WhatGAlsoDoes}
b_1 h_1 = 1 - i_1 
p_1 \, .
\end{equation}
Equation \eqref{WhatGDoes} is a rewriting of eq.~\eqref{BulkHomotopy} above if we set $h_1(\phi^*,\pi^*) = h(\phi^*)$. It remains to check \eqref{WhatGAlsoDoes}:
\begin{equation}
\begin{split}
(1 - i_1 p_1)(\phi^*,\pi^*) &= (\phi^*,p_0 \partial_n h(\phi^*)) \\
&= (Lh(\phi^*),p_0 \partial_n h(\phi^*)) = \partial h(\phi^*,\pi^*) \, . 
\end{split}
\end{equation}

The existence of the homotopy together with \eqref{PIis1} tells us that the chain map in \eqref{BoundaryCMap} is a quasi-isomorphism, i.e.~that $p_0$ and $p_1$ descend to an isomorphism on cohomology, with the inverse being  the descendants of $i_0$ and $i_1$. The bulk algebra and the boundary algebra are therefore equivalent.

We already noted that the differential $\bar \partial$ can also be derived from the action $\bar S = S\circ i_0$ due to the Hamilton-Jacobi equations. This is due to the fact that $i = (i_0,i_1)$ is a cyclic $L_\infty$ morphism (since it is linear, it is trivially so) to the theory with differential $c_1$ and symplectic form
\begin{equation}
\bar \omega(\pi^*,\varphi) = \omega(i_0(\varphi),i_1(\pi^*)) = \int_{\partial M} \text d A \, \pi^*(x) \varphi(x) \, .
\end{equation}
The boundary action $\bar S = S\circ i$ can then be written as
\begin{equation}
\bar S(\varphi) = \frac{1}{2}\bar{\omega}(c_1(\varphi),\varphi) \, .
\end{equation}

\subsection{Example: The Harmonic Oscillator}

Let us illustrate the above machinery by  the one-dimensional harmonic oscillator defined on an interval $I = [t_i,t_f]$ (for a closely related 
homological perspective on the harmonic oscillator, see \cite{Chiaffrino:2021pob}). 
The action is given by
\begin{equation}
S[x] = \int_I \text d t \frac{1}{2}(\dot x^2 - \omega^2 x^2) \, , 
\end{equation}
and the differential reads
\begin{equation}
\begin{split}
b_1: C^{\infty}(I) &\rightarrow C^{\infty}(I)\oplus C^{\infty}(\partial I) \ , \\
x &\mapsto (-\ddot x - \omega^2 x,\dot x|_{\partial I}) \, . 
\end{split}
\end{equation}
Note that $\partial I = \{t_i,t_f\}$, so $C^{\infty}(\partial I) \cong \mathbb{R}^2$, under which $x|_{\partial I}$ becomes $(x(t_i),x(t_f))$. The extended symplectic structure is
\begin{equation}
\omega((x^*,\pi_i^*,\pi_f^*),x) = \int_I \text d t \, x^*(t)x(t) + \pi_f^* x(t_f) - \pi_i^* x(t_i) \, , 
\end{equation}
with which one  correctly obtains 
\begin{equation}
S[x] = \frac{1}{2}\omega(b_1(x),x) \, .
\end{equation}

We will now give the bulk-to-bulk propagator, from which we will derive the boundary-to-bulk propagator using the general recipe we developed. Given a function $t \mapsto  x^*(t)$, the solution to $L(x) = -\ddot x -\omega^2 x = x^*$ with $x(t_i) = x(t_f) = 0$ is
\begin{equation}\label{QMPropagator}
x(t) = \int_I(K(t,s)+K(s,t))x^*(s) \ ,
\end{equation} 
where
\begin{equation}
K(t,s) = -\theta(t-s) \frac{\sin(s-t_i)\sin \omega(t-t_f)}{\sin \omega(t_f-t_i)} \ .
\end{equation}
The fact that the integral kernel in \eqref{QMPropagator} is symmetric in $t$ and $s$ implies that $h$ is symmetric with respect to the bulk symplectic structure
\begin{equation}
\omega_I(x^*,x) = \int_I \text d t \, x^*(t) x(t) \ .
\end{equation}

The inclusion map $i$ was constructed using the Hamiltonian symplectic form $\omega_\partial$, which derives from $\omega_I$ as the failure of $L$ being cyclic. We have
\begin{equation}
\omega_I(Lx_1,x_2) + \omega_I(x_1,L x_2) = (x_1 \dot x_2 - \dot x_1 x_2)|^{t_f}_{t_i} = \omega_\partial(\pi(x_1),\pi(x_2)) \ ,
\end{equation}
where
\begin{equation}
\omega_\partial((x_i,p_i,x_f,p_f),(y_i,y_f,q_i,q_f)) = x_f q_f - p_f y_f - x_i q_i + p_i y_i
\end{equation}
and\begin{equation}
\begin{split}
\pi: C^\infty(I) &\rightarrow C^\infty(\partial I) \oplus C^\infty(\partial I) \cong \mathbb{R}^4 \ , \\
x &\mapsto (x(t_f),\dot x(t_f),x(t_i), \dot x(t_i)) \ .
\end{split}
\end{equation}
The inclusion $\iota$ is then defined by
\begin{equation}
\omega_I(x^*,\iota(x_i,p_i,x_f,p_f)) = \omega_\partial(\pi h(x^*),(x_i,p_i,x_f,p_f)) \ .
\end{equation}
On the right hand side we have
\begin{equation}
\pi h(x^*) = (0,-\int_I \text d s \frac{\sin \omega(s-t_i)}{\sin \omega(t_f - t_i)}x^*(s),0,-\int_I \text d s \frac{\sin \omega(s-t_f)}{\sin \omega(t_f - t_i)}x^*(s))\,, 
\end{equation}
and hence
\begin{equation}
\begin{split}
\omega_\partial(\pi h(x^*),(x_i,p_i,x_f,p_f)) =& \int_I \text d s \, x^*(s) \frac{\sin \omega(s-t_i)}{\sin \omega(t_f - t_i)} x_f \\
&+ \int_I \text d s \, x^*(s)\frac{\sin \omega(t_f-s)}{\sin \omega(t_f - t_i)}x_i \ .
\end{split}
\end{equation}
From this we deduce that
\begin{equation}
\iota(x_i,p_i,x_f,p_f)(t) = x_f\frac{\sin \omega(t-t_i)}{\sin \omega(t_f-t_i)}+x_i\frac{\sin \omega(t_f-t)}{\sin \omega(t_f - t_i)} \ .
\end{equation}
As expected, this map is independent of the momenta $(p_i,p_f)$. In the general discussion  this observation suggested the definition of 
the improved inclusion map
\begin{equation}
i(x_i,x_f) = \iota(x_i,0,x_f,0) \ .
\end{equation}
In the other direction, we have the projector $p: x \mapsto x|_{\partial I} = (x(t_i),x(t_f))$. It satisfies $pi = 1$, which means that $i(x_i,x_f)$ is indeed the solution to $Lx = 0$ with boundary conditions $x(t_i) = x_i$ and $x(t_f) = x_f$. It is also straightforward to check that $1 - ip = hL$.

The extended homotopy data that includes the boundary part is easy to read off from our general discussion. We have
\begin{align}
&p_0(x) = (x(t_i),x(t_f)) & \ ,\\ 
&p_1(x^*,\pi_i^*,\pi_f^*) = (\pi_i^* + \int_I \text d s \frac{\sin \omega(s-t_i)}{\sin \omega(t_f - t_i)}x^*(s),\pi_f^* + \int_I \text d s \frac{\sin \omega(s-t_f)}{\sin \omega(t_f - t_i)}x^*(s)) &  \ , \\ 
&i_0(x_i,x_f) = i(x_i,x_f) \ , \quad  i_1(\pi_i^*,\pi_f^*) = (0,\pi_i^*,\pi_f^*) \ .
\end{align}
The boundary action then is
\begin{equation}
S_\partial[x_i,x_f] = \frac{\omega(x_f^2 + x_i^2)}{2}\frac{\cos \omega(t_f - t_i)}{\sin\omega(t_f - t_i)} - \frac{\omega x_i x_f}{\sin\omega(t_f - t_i)} \ .
\end{equation}
Just to make sure that this is correct, we compute $\partial_{x_i} S$ and $\partial_{x_f} S$ to check the Hamilton-Jacobi equations: 
\begin{equation}
\begin{split}
\partial_{x_f} S &= \frac{\omega x_f \cos \omega(t_f - t_i)- \omega x_i}{\sin \omega(t_f - t_i)} = \dot x(t_f) \ , \\
\partial_{x_i} S &= \frac{\omega x_i \cos \omega(t_f - t_i)- \omega x_f}{\sin \omega(t_f - t_i)} = -\dot x(t_i) \ , 
\end{split}
\end{equation}
which are indeed satisfied.

\section{Homotopy Retract for Gauge Theories} 

In this section we apply the ideas we developed above for the scalar field to Yang-Mills theory. To this end we review the Hodge decomposition for manifolds 
with boundary, which is more involved compared to the familiar Hodge decompositions of manifolds without boundary.

\subsection{Hodge Decomposition on Manifolds with Boundary}

The following review is based on \cite{Schwarz1995HodgeD}. Let $M$ be a $d$-dimensional compact Riemannian manifold with metric $g$. We denote by $\Omega^{k}(M)$ the space of $k$-forms on $M$. The metric allows us to define a Hodge star operator $ \star: \Omega^k(M) \rightarrow \Omega^{d-k}(M) 
$, satisfying  $\star^2 = (-)^{k(d-k)}$ on $k$-forms,  
which defines an inner product on $k$-forms for any $k=0,\ldots, d$: 
\begin{equation}\label{YMinnerproduct}
\begin{split}
\Omega^k(M)\otimes \Omega^k(M) \longrightarrow \mathbb{R} \, , \qquad 
(a,b) := \int_M a \wedge \star b \, .
\end{split}
\end{equation}

Let $\text d: \Omega^k(M) \rightarrow \Omega^{k+1}(M)$ be the de-Rham differential. We define its adjoint 
via $\delta = -(-)^{n(n-k)}\star\text d \, \star$, 
with $k$  the degree of the form $\delta$ acts on. This  is used to define the Laplacian $\Delta = \text d \delta + \delta \text d$.
When $M$ does not have a boundary, $\delta$ is the adjoint of $\text d$ with respect to the inner product $(-,-)$, 
but in presence of a boundary there are corrections in the form of boundary terms. 
In order to determine these boundary terms let us introduce, for a 
given manifold with boundary,  a vector field $N$ normal to the boundary, which we normalize  such that $g(N,N)|_{\partial M} = 1$. We denote by $\Gamma(TM)|_{\partial M}$ the space of sections of the tangent bundle $TM$ restricted to the boundary. Note that this is not $\Gamma(T\partial M)$, since sections in $\Gamma(TM)|_{\partial M}$ may not be tangent to $\partial M$. The normal field $N$ then defines a splitting
\begin{equation}
\Gamma(TM)|_{\partial M} = \Gamma(TM)|_{\partial M}^{\parallel} \oplus \Gamma(TM)|_{\partial M}^{\perp} \, .
\end{equation}
Here, sections in $\Gamma(TM)|_{\partial M}^{\parallel}$ are those proportional to $N$, while the sections in $\Gamma(TM)|_{\partial M}^{\perp}$ include  those orthogonal to $N$. We have $\Gamma(TM)|_{\partial M}^{\perp} \cong \Gamma(T\partial M)$.

Given a $k$-form $\alpha$, we write $\mathbf t\alpha$ for its restriction to $\Gamma(TM)|_{\partial M}^{\perp}$. Since the direct sum decomposition allows us to define a projector $\Gamma (TM)|_{\partial M} \rightarrow \Gamma(TM)|_{\partial M}^{\perp}$, we can think of $\mathbf t$ as an endomorphism on $\Gamma (TM)|_{\partial M}$. This in turn allows us to define $\mathbf n \alpha = \alpha|_{\Gamma(M)|_{\partial M}} - \mathbf t \alpha$. We then write $\Omega_D^k(M) = \ker \mathbf t$ and $\Omega_N^k(M) = \ker \mathbf n$, where the subscripts stand for Dirichlet and Neumann boundary conditions. To make contact with the previous section, note that the scalar fields in $\Omega_D^0(M)$ are exactly the scalar fields vanishing on the boundary. 

We can use the maps $\mathbf t$ and $\mathbf n$ to give a version of Green's theorem for forms. The failure of $\delta$ to be the adjoint of $\text d$ is a boundary term,
\begin{equation}
(\text d \alpha,\beta) - (\alpha,\delta \beta) = \int_{\partial M} \mathbf t \alpha \wedge \star \mathbf n \beta \ .
\end{equation}
The failure obviously vanishes if either $\alpha$ satisfies Dirichlet or $\beta$ satisfies Neumann boundary conditions.

We now come to the Hodge decomposition in the presence of a boundary. We define the space of harmonic $k$-forms to be $\mathcal{H}^k := \ker \text d|_{\Omega^k(M)} \cap \ker \delta|_{\Omega^k(M)}$. We then have a decomposition (theorem 2.4.2 in \cite{Schwarz1995HodgeD})
\begin{equation}\label{HodgeDecomp}
\Omega^k(M) = \text d\Omega^{k-1}_D(M) \oplus \delta\Omega^{k+1}_N(M) \oplus \mathcal{H}^k(M) \ .
\end{equation}
This decomposition is orthogonal with respect to $(-,-)$. Also, $\text d$ (respectively $\delta$) preserves Dirichlet (respectively  Neumann) boundary conditions, so forms in $\text d\Omega^{k-1}_D(M)$ (respectively  in $\delta \Omega^{k+1}_N(M)$) satisfy the same boundary conditions.

Following theorem 2.4.8 in 
 \cite{Schwarz1995HodgeD}, the decomposition can be further improved in two distinct ways. We define $\mathcal{H}^k_{ex}(M) := \text d \Omega^{k-1}(M) \cap \mathcal H^k(M)$ and $\mathcal{H}^k_{coex}(M) := \delta \Omega^{k+1}(M) \cap \mathcal H^k(M)$. We further denote by $\mathcal H_D^k(M)$ and $\mathcal H_N^k(M)$ the spaces of harmonic forms with boundary conditions indicated by the subscripts. We then have
\begin{equation}
\mathcal{H}^k(M) = \mathcal H_D^k(M)\oplus \mathcal{H}^k_{coex}(M) = \mathcal H_N^k(M)\oplus \mathcal{H}^k_{ex}(M) \ .
\end{equation}
Both decompositions are again orthogonal with respect to $(-,-)$.

We end the discussion on the Hodge decomposition by giving an interpretation in terms of de-Rham cohomology. This will later help to understand the form of topological solutions in Yang-Mills theory. However, it is not necessary to follow the rest of the paper. 

Without a boundary, the de-Rham cohomology can be identified with the harmonic forms. This is no longer true here. To compute the cohomology, we decompose a given $k$-form $\alpha$ using \eqref{HodgeDecomp}. We write
\begin{equation}
\alpha = \text d a + \delta b + c \ ,
\end{equation}
where $a$ satisfies Dirichlet boundary conditions, $b$ satisfies Neumann boundary conditions and $c$ is harmonic. We find that $\text d \alpha = 0$ if and only if $\delta b = 0$, since $\text d \alpha = \text d \delta b$ and
\begin{equation}
(\text d\delta b,b) = (\delta b,\delta b) \ge 0 \ .
\end{equation} 
Here we used Green's theorem, where the boundary term is zero since $b$ satisfies Neumann boundary conditions. The inner product $(-,-)$ is positive definite, so $\text d \alpha = \text d \delta b = 0$ implies $\delta b = 0$. Hence, any closed form is given by
\begin{equation}
\alpha = \text d a + c \ . 
\end{equation}
We further write $c = e + \text d f \in \mathcal{H}^k_N(M) \oplus \mathcal{H}_{ex}^k(M)$. Since this decomposition is a direct sum, $e$ cannot be exact unless it is zero. Therefore $\alpha = e$ up to exact elements. So we can identify the $k$th de-Rham cohomology with $\mathcal{H}^k_N(M)$.

The space $\mathcal{H}_D^k(M)$ has a similar interpretation. Since the pullback of any function commutes with the exterior derivative, we can restrict $\text d$ to the kernel $\ker f^*$ of any smooth map $f$. In particular, we can restrict it to $\ker j^*$, where $j: \partial M \rightarrow M$ is the boundary inclusion. As noted before, we have $\Omega_D^\bullet(M) = \ker j^*$. The cohomology of $\ker j^*$ is often denoted by $H^\bullet(M,\partial M)$ 
and  called the relative de-Rham cohomology with respect to $\partial M$. We will show that $H^k(M,\partial M) \cong \mathcal{H}_D^k(M)$. For any $k$-form $\alpha$, we again use the de-Rham composition
\begin{equation}
\alpha = \text d a + \delta b + c
\end{equation}
as before and further decompose $c = e + \delta f \in \mathcal{H}_D^k(M) \oplus \mathcal{H}_{coex}^k(M)$. We already learned that $\text d \alpha = 0$ implies $\delta b = 0$. We now assume that $\alpha$ further satisfies Dirichlet boundary conditions. $\text d a$ and $e$ satisfy this condition by assumption. On the other hand, $\delta f$ cannot satisfy Dirichlet boundary conditions unless it is zero, since $\mathcal{H}_D^k(M) \cap \mathcal{H}_{coex}^k(M) = \{0\}$ due to the fact that the decomposition of $c$ is a direct sum. This proves that a closed form $\alpha$ satisfying Dirichlet boundary conditions can be written as $\alpha = e \in \mathcal{H}^k_D(M)$ up to exact terms. This proves the claim $H^k(M,\partial M) \cong \mathcal{H}_D^k(M)$.

\subsection{Yang-Mills Theory as a cyclic $L_\infty$ Algebra}

\label{sec:YMproducts}

\textit{Cyclic $L_{\infty}$ Algebra of Yang-Mills Theory without a Boundary}\\[-3ex]

We now describe Yang-Mills as an $L_\infty$ algebra, which was previously demonstrated in \cite{Zeitlin:2007ttl,Zeitlin:2008cc,Zeitlin:2009zz}. We fix an $n$-dimensional manifold $M$. 
Yang-Mills fields are then given by Lie algebra valued one-forms on $M$,  
at least for the sake of perturbation theory. Under the same assumption, the gauge field $A$ globally splits into a form component and a Lie algebra component, i.e.~$A \in \Omega^1(M)\otimes \mathfrak{g}$ for some fixed Lie algebra $\mathfrak{g}$.

The curvature of the gauge field $A \in \Omega^1(M)\otimes \mathfrak{g}$ is given by $F = \text d A + \frac{1}{2}[A,A]$, where $[-,-]$ combines the wedge product of forms with the Lie bracket of the Lie algebra. The Yang-Mills action reads 
\begin{equation}
S_0[A] = \frac{1}{2}\int_M \tr ( F \wedge \star F) \, , 
\end{equation}
and its Euler-Lagrange equations are  given by
\begin{equation}
0 = \text d A + \frac{1}{2}\text d\star[A,A] + [A,\star \text d A] + \frac{1}{2}[A,\star [A,A]] \ .
\end{equation}

The chain complex describing the free part of  Yang-Mills theory is given by
\begin{equation}
\begin{tikzcd}
\Omega^0(M)\otimes \mathfrak{g} \arrow[r,"\text d"] & \Omega^1(M)\otimes \mathfrak{g} \arrow[r,"\text d \star\text d"] & \Omega^{n-1}(M)\otimes \mathfrak g \arrow[r,"\text d"] & \Omega^n(M)\otimes \mathfrak{g} \ .
\end{tikzcd} 
\end{equation}
Here $\text d$ is the usual de-Rham differential acting trivially on the Lie algebra part. Since we want the physical fields to live in degree zero, the complex starts in degree $-1$ and ends in degree $2$. As usual, we denote the differential collectively by $b_1$. From now on, we will write $\Omega^\bullet(M)$ for $\Omega^\bullet(M) \otimes \mathfrak g$.

Let us find the $L_\infty$ products. Below, any product not explicitly given is zero. 
 The $L_\infty$ products should be graded symmetric and such that
\begin{equation}
b_1(A) + \frac{1}{2}b_2(A,A) + \frac{1}{6}b_3(A,A,A) = 0
\end{equation}
is the equation of motion. The products on fields are then
\begin{equation}
b_2(A_1,A_2) = \text d \star [A_1,A_2] + [A_1, \star \text d A_2] + [A_2,\star \text d A_1] \ ,
\end{equation}
and 
\begin{equation}
b_3(A_1,A_2,A_3) = [A_1,\star [A_2,A_3]] + [A_2,\star [A_3,A_1]] + [A_3,\star [A_1,A_2]] \ .
\end{equation}
The products on gauge parameters $c \in \Omega^{0}(M)$ are such that
\begin{equation}
\delta_c A = b_1(c) + b_2(c,A)
\end{equation}
are the gauge transformations of $A$. Since these are given by $\delta A = \text d c + [A,c]$, we need to define
\begin{equation}
b_2(c,A) = b_2(A,c) = -[c,A]
\end{equation}
for any gauge parameter $c$ and any gauge field $A$. For this bracket to satisfy the Jacobi identity, we further need
\begin{equation}
b_2(c_1,c_2) = -[c_1,c_2]
\end{equation}
for any two gauge parameters $c_1,c_2$.
Gauge invariance of the theory is ensured by gauge covariance of the equations of motion. Let $A^* \in \Omega^{d-1}(M)$, 
i.e.~$A^*$ lives in the space of equations of motion. The gauge transformations of $A^*$ are encoded in
\begin{equation}
b_2(c,A^*) = - b_2(A^*,c) = -[c,A^*] \ .
\end{equation}

One can show that the products defined above satisfy the $L_\infty$ relations if we ignore the space $\Omega^d(M)\otimes \mathfrak{g}$ in degree 2. This space is needed, however, in order to later define a cyclic structure. We call the vector space $\Omega^d(M)\otimes \mathfrak{g}$ the space of Noether identities, and denote its elements or  fields  by $c^*$. We can then extend the products via
\begin{equation}
b_2(A,A^*) = b_2(A^*,A) = [A,A^*]
\end{equation}
for any gauge field $A$ and any equation of motion $A^*$, as well as
\begin{equation}
b_2(c,c^*) = b_2(c^*,c) = -[c,c^*]
\end{equation}
for any gauge parameter $c$ and any Noether identity $c^*$. These products reflect the fact that
\begin{equation}
\text D^2_A \star F = 0 \ .
\end{equation}

Without a boundary, the cyclic structure of Yang-Mills theory can be defined by
\begin{equation}
\omega(c,c^*) = -\omega(c^*,c) = \int_M c \wedge c^* \ , \qquad \omega(A^*,A) = - \omega(A,A^*) = \int_M A \wedge A^* \ .
\end{equation}
Recall from \ref{CyclicGeneralities} that we can extend the classical action $S_0$ to an action $S$, defined as a multilinear function
\begin{equation}
S = \sum_{n \ge 1}\omega(b_n\otimes 1) \ .
\end{equation}
In the case of Yang-Mills theory, we find
\begin{equation}\label{BVYangMills}
\begin{split}
S &= \int_M \frac{1}{2} A \wedge \Big(\text d \star \text d A + \frac{1}{2}\text d \star [A,A] + [A,\star \text d A]+ \frac{1}{4}[A,[A,A]]\Big)  \\
& \quad - \int_M (\text d c + [c,A])\wedge A^* + \frac{1}{2} c^* \wedge [c,c] \\
&= \frac{1}{2}  \int_M F \wedge \star F + A^*\wedge D_A c + \frac{1}{2}c^*\wedge[c,c] \ .
\end{split}
\end{equation}
This is the BV extended action of Yang-Mills theory. In our case, the above expression is understood as a multilinear map as follows. The variables $\{c^*,A^*,A,c\}$ are not the fields of the complex, but rather define projection maps to the subspaces the fields live in. For example
\begin{equation}
c^*: (\Omega^0(M) \oplus \Omega^1(M) \oplus \Omega^{d-1}(M)\oplus \Omega^d(M)) \rightarrow \Omega^d(M) \ .  
\end{equation}
A wedge product of $n$ field variables $\alpha_1,...,\alpha_n \in \{c^*,A^*,A,c\}$ on $n$ fields $\mathcal A_1,...,\mathcal A_n$ in the complex is then
\begin{equation}
(\alpha_1 \wedge ... \wedge \alpha_n)(\mathcal{A}_1,...,\mathcal{A}_n) = \sum_{\sigma \in \mathcal{S}_n} \pm \alpha_1(\mathcal{A}_{\sigma(1)}) \wedge ... \wedge \alpha_n(\mathcal{A}_{\sigma(n)}) \ .
\end{equation}
The wedge product on the right is the wedge product on forms combined with the product of Lie algebra elements. The sign is the natural one coming from the degree of the fields (which, we recall,  in general is not the form degree).\footnote{Another way of thinking of the wedge product in the action is the following. Suppose we have multilinear maps $f,g: S^c(V^\bullet) \rightarrow \Omega^\bullet(M)$, where $V^\bullet$ is the underlying complex of the theory. The domain $S^c(V)$ has a coproduct $\Delta$, while the image $\Omega^\bullet(M)$ has a product $\nabla$. We can then define an associative product $f\wedge g := \nabla (f \otimes g)\Delta$ on the space of functions from $S^c(V)$ to $\Omega^\bullet(M)$. It is straightforward to see that this agrees with the definition given in the text and also makes it manifest that the product is associative.} With this it is straightforward to check that \eqref{BVYangMills} is exactly $S = \sum_{n \ge 1}\omega(b_n \otimes 1)$.\\[2ex]
\bigskip 
\textit{Cyclic $L_{\infty}$  Algebra in Presence of Boundary}\\[-1.5ex]
When the manifold has a boundary, cyclicity of the differential $b_1$ with respect to $\omega$ fails due to boundary terms, just like in the case of the scalar field. We find
\begin{equation}\label{Acycfailure}
\omega(b_1 A_1,A_2) + \omega(A_1,b_1 A_2) = \int_{\partial M} A_1 \wedge \star \text d A_2 - \int_{\partial M} A_2 \wedge \star \text d A_1 = \omega_{\partial}(A_1,A_2) \ .
\end{equation} 
The failure defines a symplectic structure, pairing the gauge field $A$ with its canonical momentum $\star \text d A$. There is further an induced symplectic structure due to the gauge symmetry:
\begin{equation}\label{ccycfailure}
\omega(b_1 c, A^*) - \omega(c, b_1 A^*) = - \int_{\partial M} c \wedge A^* = \omega_{\partial}(c,A^*)\ .
\end{equation}

To define the symplectic structure independently of the bulk data, we introduce the following spaces,
\begin{equation}
X^{-1} = \Omega^{0}(\partial M) \ , \qquad X^0 = \Omega^1(\partial M) \times \Omega^{n-2}(\partial M) \ , \qquad X^1 = \Omega^{n-1}(\partial M) \ .
\end{equation}
We can then define the symplectic structure using the obvious pairing. Further, there exists a projection from the bulk data to the boundary data via
\begin{equation}
\begin{split}
\pi: V^\bullet &\rightarrow X^\bullet = \Omega^{0}(\partial M) \oplus \Omega^1(\partial M) \times \Omega^{n-2}(\partial M) \oplus\Omega^{n-1}(\partial M)\; , \\
(c,A,A^*,c^*) &\mapsto (i^* c, i^* A, i^* \star \text d A, i^* A^*)  \; .
\end{split}
\end{equation}
Here, $i: \partial M \rightarrow M$ is the inclusion and $i^*$ its pullback. The failure of $b_1$ being cyclic can now be written as
\begin{equation}
\omega(b_1\mathcal{A},\mathcal{B}) + (-)^{\mathcal{A}}\omega(\mathcal{A},b_1 \mathcal{B}) = \omega_\partial(\pi \mathcal{A},\pi \mathcal{B}) \ ,
\end{equation}
which, like in the scalar field case, is the $L_\infty$ version of the BV-BFV relation \cite{Cattaneo:2012qu}. 

To ensure cyclicity of the theory, we introduce the following complex
\begin{equation}
\begin{tikzcd}
\Omega^0(M) \arrow[r,"b_1"] & \Omega^1(M) \arrow[r,"b_1"] & \Omega^{n-1}(M) \oplus \Omega^{n-2}(\partial M) \arrow[r,"b_1"] & \Omega^n(M)\oplus \Omega^{n-1}(\partial M) \ ,
\end{tikzcd}
\end{equation}
where now
\begin{equation}
b_1(c) = \text d c \ , \qquad b_1(A) = (\text d \star \text d A,i^*\star \text d A) \ , \qquad b_1(A^*,B^*) = (\text d A^*,i^* A^* - \text d B^*) \ .
\end{equation} 
We also extend the inner product via
\begin{equation}
\begin{split}
\omega(c,(c^*,\eta^*)) &= \int_M c \wedge c^* - \int_{\partial M} i^* c \wedge \eta^* \; , \\
\omega((A^*,B^*),A) &= \int_M A \wedge A^* + \int_{\partial M} i^* A \wedge B^* \ ,
\end{split}
\end{equation}
which makes the free theory cyclic.

Since the cubic interaction $b_2$ has derivatives, it picks up boundary terms. To ensure cyclicity, we need to extend it as
\begin{align}
b_2'(A_1,A_2) = (b_2(A_1,A_2), i^*\star[A_1,A_2]) \ , & & b_2'(A,(A^*,B^*)) = (b_2(A,A^*),-[i^* A, B^*]) \ , \\
b_2'((A^*,B^*),c) = (-[A^*,c], -[B^*,i^* c]) \ , & & b_2'((c^*,\eta^*),c) = (b_2(c^*,c),[\eta^*,i^*c]) \ .
\end{align}
For the three-bracket, it turns out that $b_3$ does not need any corrections, i.e.~it is still non-zero on fields only and it extends from $\Omega^{n-1}(M)$ to $\Omega^{n-1}(M)\times \Omega^{n-2}(\partial M)$ by zero: 
\begin{equation}
b_3'(A_1,A_2,A_3) = (b_3(A_1,A_2,A_3),0) \ .
\end{equation}

\subsection{A Homotopy for Yang-Mills Theory}

\label{Sec:YMHomotopy}

We will now give a homotopy, first for the non-cyclic variant of Yang-Mills theory. We give a more in-depth derivation of the homotopy in appendix \ref{App:YMHomotopy}. As we did with the scalar field, we then extend it to the cyclic theory.

We first give the propagator, i.e.~the homotopy on equations of motion $A^* \in \Omega^{n-1}(M)$. Let $P_1: \Omega^{n-1}(M) \rightarrow \mathcal{H}_N^{n-1}$ be the orthogonal projection to $\mathcal{H}_N^{n-1}$ and $I_1: \mathcal{H}_N^{n-1} \rightarrow \Omega^{n-1}(M)$ be the canonical inclusion. Then, let $G_1$ be the unique Green's function solving
\begin{equation}
\Delta G_1(A^*) = \star(1 - I_1 P_1)A^* \; ,
\end{equation}
such that $G_1(A^*) \in (\mathcal{H}_D^1(M))^\perp$. We also introduce the orthogonal projection $P_2: \Omega^{n-1}(M) \rightarrow \delta \Omega^n_N(M)$ and the canonical inclusion $I_2: \delta \Omega^{n} \rightarrow \Omega^{n-1}(M)$. 
We define the homotopy on $A^*$ to be
\begin{equation}
h_1(A^*) = G_1(1-I_2P_2)(A^*) \; .
\end{equation}
We show in the appendix that $h$ satisfies Lorenz gauge, i.e.~$\delta h = 0$.

Let $G_0: \Omega^0(M) \rightarrow \Omega^0_D(M)$ be the scalar Green's function we are familiar with from our discussion of the scalar field. On gauge fields $A$, we use it to define the homotopy
\begin{equation}
h_0(A) = G_0 \delta A \, .
\end{equation}
We then find
\begin{equation}
h_0b_1 c = G_0 \Delta c = (1 - i_{-1}p_{-1})(c) \; ,
\end{equation}
where $p_{-1} = i^*$, i.e.~it restricts the gauge parameter to the boundary and $i_{-1}$ computes the harmonic $c$ with given boundary value.

With the homotopy on fields and equations of motion at hand, we can now compute the commutator $\{b_1,h\}$. We have
\begin{equation}
(1-i_0p_0)(A) = (b_1 h_0 + h_1 b_1)(A) \; ,
\end{equation}
where $p_0: \Omega^1(M) \rightarrow \mathcal{H}_D^1(M) \oplus \Omega^1(\partial M)$ is the projection defined by $p_0(A) = (P_3(A),i^* A)$, with $P_3: \Omega^1(M) \rightarrow \mathcal{H}_D^1(M)$ the orthogonal projection. In the other direction, let $u: \Omega^1(\partial M) \rightarrow \Omega^1(M)$ be the map that associates to a one form $a_0 \in \Omega^1(\partial M)$ the unique solution $u(\alpha) \in \Omega^1(M)$ satisfying $\Delta u(a_0) = 0$, such that $i^* r(a_0) = a_0, i^* \delta u(a_0) = 0, u(a_0) \in (\mathcal{H}_D^1)^\perp(M)$. Further, let $I_3: \mathcal H^1_D(M) \rightarrow \Omega^1(M)$ be the canonical inclusion. We then define $i_0: \mathcal{H}^1_D(M) \oplus \Omega^1(\partial M) \rightarrow \Omega^1(M)$ as the sum $i_0 = I_3 + r$.

The last piece of the homotopy is given by $h_2: \Omega^n(M) \rightarrow \Omega^{n-1}(M)$, where 
 \be
   h_2 = (-)^n \delta \star G_0 \star\,. 
 \ee  
   One can show that $b_1 h_2 = 1$. This is consistent with the fact that any connected compact manifold with boundary has no cohomology in top degree. On the other hand, $h_2 b_1 + b_1 h_1 = 1 - i_1 p_1$, where $p_1 = P_1$ and $i_1 = I_1$.

The homotopy $h_\bullet$, together with the chain maps $i_\bullet$ and $b_\bullet$ give rise to a homotopy
\begin{equation}
\begin{tikzcd}
X^\bullet: 0 \arrow[r] & \Omega^0(M) \arrow[r,"\text d"] \arrow[d,"p_{-1}"] & \Omega^1(M) \arrow[r,"\text d \star \text d"] \arrow[d,"p_0"] & \Omega^{n-1}(M) \arrow[r,"\text d"] \arrow[d,"p_1"] & \Omega^n(M) \arrow[r] \arrow[d,"0"] & 0 \\
Y^\bullet: 0 \arrow[r] & \Omega^0(\partial M) \arrow[r,"\text (0{,}\text d)"] &  \mathcal{H}_D^1(M) \oplus \Omega^1(\partial M) \arrow[r,"0"] & \mathcal{H}_N^{n-1}(M) \arrow[r,"0"] & 0 \arrow[r] & 0 \, .
\end{tikzcd}
\end{equation}
As in the scalar field case, the complex $Y^\bullet$ of the boundary theory is not symmetric, in the sense that $Y^i$ is not isomorphic to $Y^{1-i}$. This just reflects the fact that the theory is not cyclic. We will now extend the homotopy to the cyclic complex, which we introduced in the previous section.

Recall that in order to obtain a cyclic theory, we extended both the space of equations of motion and the space of Noether identities in the following way:
\begin{equation}
\Omega^{n-1}(M) \mapsto \Omega^{n-1}(M) \times \Omega^{n-2}(\partial M) \ , \qquad \Omega^n(M) \mapsto \Omega^n(M)\times \Omega^{n-1}(\partial M) \ .
\end{equation}
For the boundary complex, we make the ansatz
\begin{equation}
\begin{tikzcd}
\Omega^0(\partial M) \arrow[r,"c_1"] & \Omega^1(\partial M) \oplus \mathcal{H}_D^1(M) \arrow[r,"c_1"] & \Omega^{n-2}(\partial M) \oplus \mathcal{H}_N^{n-1}(M) \arrow[r,"c_1"] & \Omega^{n-1}(\partial M) \ .
\end{tikzcd}
\end{equation}
We denote the boundary field by $\eta \in \Omega^0(\partial M)$, $(a_0,\mathcal{A}) \in \Omega^1(\partial M) \times \mathcal{H}_D^1(M)$, $(B^*,\mathcal{A}^*)$ and $\eta^* \in \Omega^{d-1}(\partial M)$. Here, we define
\begin{equation}
c_1(c) = (\text d c,0) \ , \qquad c_1(a_0,\mathcal A) = (i^*\star\text d i_0(a_0, \mathcal A),0)\ , \qquad c_1(B^*,\mathcal A^*) = \text d B^* - i^*i_1(\mathcal A^*) \ .
\end{equation}
Note that, among other reasons, $c_1^2 = 0$ follows from $\text d \star \text d i_0(B,\mathcal A) = 0$, since the image of $i_0$ are solutions to the equations of motion. Also, we have been particularly pedantic when using the inclusion $i_1: \mathcal{H}_N^{n-1}(M) \rightarrow \Omega^{n-1}(M)$ before applying the pullback $i^*$.

As an ansatz for the homotopy, we will extend $h_1$ trivially to a map
\begin{equation}
h_1': \Omega^{n-1}(M) \oplus \Omega^{n-2}(\partial M) \longrightarrow \Omega^1(M) \ ,
\end{equation}
i.e.~we set $h_1'(A^*,B^*) = h_1(A)$, where on the right hand side  we denote by $h_1$ the original homotopy. In this way, we do not affect the action of the homotopy on the gauge fields. On the other hand, for an equation of motion $(A^*,B^*) \in X^1$, we find
\begin{equation}
b_1' h_1'(A^*,B^*) = (b_1 h_1(A^*), i^*\star \text d h_1(A^*)) \ .
\end{equation}
The second component is the canonical momentum of $h_1(A^*)$. We also need an extension $h_2'$ of $h_2$, 
which can be obtained by defining
\begin{equation}
\begin{split}
h_2': \Omega^n(M) \times \Omega^{n-1}(\partial M) &\longrightarrow \Omega^{n-1}(M) \times \Omega^{n-2}(\partial M) \\
(c^*,\eta^*) &\longmapsto (h_2(c^*),0) \ .
\end{split}
\end{equation}
We then find
\begin{equation}
(b_1' h_1' + h_2' b_1')(A^*,B^*) = (A^*,B^*) - (i_1 p_1(A^*),B^* - i^* \star \text d h_1(A^*)) \ .
\end{equation}
Furthermore,
\begin{equation}
b_1' h_2'(c^*,\eta^*) = (c^*,\eta^*) - (0,\eta^* -  i^* h_2 (c^*)) \ .
\end{equation}
This suggests that we define the following projections:
\begin{equation}
\begin{split}
p_1': \Omega^{n-1}(M)\times \Omega^{n-2}(\partial M) &\longrightarrow \Omega^{n-2}(\partial M) \times \mathcal{H}_N^{n-1}(M) \ , \\
(A^*,B^*) &\longmapsto (B^* - i^* \star \text d h_1(A^*),p_1(A^*)) \ , \\
p_2': \Omega^n(M) \times \Omega^{n-1}(\partial M) &\longrightarrow \Omega^{n-2}(\partial M) \ , \\
(c^*,\eta^*) &\longmapsto \eta^* - i^* h_2(c^*) \ .
\end{split}
\end{equation}
We also set $p_0' = p_0$ and $p_{-1}' = p_{-1}$. We need to check that these satisfy the chain map property, i.e.
\begin{equation}
c_1 p_{-1}' = p_0' b_1' \ , \qquad c_1 p_0' = p_1' b_1' \ , \qquad c_1 p_1' = p_2' b_1' \ .  
\end{equation}
From the previous section we know that the first equation holds, since the maps involved did not receive 
a cyclic extension. Now let $A \in \Omega^1(M)$. We then find
\begin{equation}
\begin{split}
p_1' b_1'(A) 	&= p_1'(i^*\star \text d A,b_1 A)= (i^* \star \text d A - i^* \star \text d h_1b_1(A),p_1b_1(A)) \\
				&= (i^* \star\text d (1-h_1 b_1)(A),0) = (i^* \star \text d (b_1 h_0 + i_0p_0)(A),0) = (i^* \star \text d i_0p_0(A),0)  \\
				&= c_1 p_0'(A) \ . 
\end{split}
\end{equation}
This proves the first chain map relation. Here, we used the homotopy relation $h_1 b_1 + b_1 h_0 = 1 - i_0 p_0$ as well as the fact that $b_1$ on gauge parameters $h_0(A)$ acts as $\text d h_0(A)$. 

Next, we consider  $(A^*,B^*) \in \Omega^{n-1}(M) \times \Omega^{n-2}(\partial M)$ and compute 
\begin{equation}
\begin{split}
c_1 p_1'(A^*,B^*) 	&= c_1(B^* - i^* \star \text d h_1 (A^*),p_1(A^*)) = \text d (B^*) - \text d i^* \star \text d h_1 (A^*) - i^*i_1 p_1(A^*)\\
					&=  \text d B^* - i^* (b_1 h_1+i_1p_1)(A^*) = \text d B^* - i^*(1 - h_2 b_1)(A^*) \\
					&= p_2'(\text d A^*, B^* - i^* A^*) 
					= p_2' b_1'(A^*,B^*)  \ .
\end{split}
\end{equation}
This completes the prove that the $p_j'$ combine into a chain map.

We also need the inclusion maps, which themselves should give rise to chain maps. In positive degree, we define them via the trivial inclusions, which are
\begin{equation}
\begin{split}
i_1': \mathcal{H}_N^{n-1}(M) \times \Omega^{n-2}(\partial M) &\rightarrow \Omega^{n-1}(M) \times \Omega^{n-2}(\partial M) \ , \\
i_2': \Omega^{n-1}(\partial M) &\rightarrow \Omega^n(M) \times \Omega^{n-1}(\partial M) \ .
\end{split}
\end{equation}
Here, $i_1$ identifies $\mathcal{H}_N^{n-1}(M)$ as a subspace of $\Omega^{n-1}(M)$ and acts on $\Omega^{n-2}(\partial M)$ as the identity, while  $i_2$ is just the inclusion into the second factor.

We further define $i_0' := i_0$ and $i_{-1}' := i_{-1}$. To prove the chain map property, we need to show that
\begin{equation}
c_1 i_{-1}' = i_0' b_1' \ , \qquad  i_1' c_1 = b_1' i_0' \ , \qquad i_2' c_1 = b_1' i_1' \ .
\end{equation}
Again, we proved the first equation already in the previous section, since all the maps involved did not  need a cyclic extension. We further have
\begin{equation}
i_1' c_1(\mathcal{A},B) = i_1'(0,i^*\star \text d i_0 B) = b_1 i_0(\mathcal{A},B) \ ,
\end{equation}
as well as
\begin{equation}
i_2' c_1(\mathcal{A}^*,B^*) = i_2'(\text d B^* - i^* i_1 \mathcal{A}^*) = (0,\text d B^* - i^* i_1 \mathcal{A}^*) = b_1' i_1'(\mathcal{A}^*,B^*) \ .
\end{equation}
This proves that the $i_j'$ combine into a chain map.

It is straightforward to see that $1- i'_i p'_i = b_1' h_i + h_{i+1}b_1'$, when we recall that we already computed the right hand side of that equation. We also have $p_i i_i = 1$, as well as $p_ih_{i+1} = 0$ and $h_i i_i = 0$.

\section{Homotopy Retract to Boundary of Anti-de-Sitter}

In this section we apply the above general framework of homotopy algebras for field theories defined on manifolds with a boundary 
to AdS/CFT. This case entails further subtleties as  AdS   does not have a boundary in the strict   topological sense, 
but only a `conformal boundary' \cite{Witten:1998qj}. 
This in turn implies that the projection operator involves a rescaling by the $z$ coordinate in the Poincar\'e patch. 
We establish the homotopy retract for this case and show that determining the homotopy transferred cyclic $L_{\infty}$ algebra, 
and hence the on-shell action, by means of  the homological perturbation lemma, amounts to the familiar method of Witten diagrams, 
hence providing a homological interpretation of the latter.

 \subsection{Homotopy Retract for AdS}

We begin by collecting some generalities about Anti-de-Sitter (AdS) space,  
which we here take to be of Euclidean signature. 
We work in the Poincar\'e patch,  
for which the metric $g$ of $(d+1)$-dimensional Euclidean $AdS$ is given by 
\begin{equation}\label{AdSmetric} 
ds^2 = \frac{1}{z^2} \Big(dz^2 + \sum_{i=1}^d dx_i^2\Big)\;. 
\end{equation}
In these coordinates the conformal boundary is given by $d$-dimensional flat (Euclidean) space with metric $\sum_{i=1}^d dx_i^2$, 
which we can think of as being obtained from (\ref{AdSmetric}) by taking   the limit $z \to 0$. 
[For instance, we can rescale $z\rightarrow \lambda z$ and then send $\lambda\rightarrow 0$, for which the second term in 
(\ref{AdSmetric}) dominates over the first term.] 

Next, we inspect a simple field theory defined on AdS, with  a free scalar field $\phi$, whose coordinate
dependence  we display, in line with (\ref{AdSmetric}),  as $\phi(x,z)$. The free action for a massive scalar then reads 
\begin{equation}
\begin{split}
S &= \frac{1}{2} \int_{AdS}  \sqrt{g} \left(g^{\mu \nu} \partial_{\mu} \phi \partial_{\nu} \phi + m^2 \phi^2\right) \\
&=  \frac{1}{2} \int \text   d^dx \text dz \left( {z^{-d+1}} \left((\partial_z\phi)^2 + (\partial_i\phi)^2\right) + {z^{-d-1}} m^2\phi^2\right) \;, 
\end{split} 
\end{equation}
where the second line follows with (\ref{AdSmetric}). 
The equations of motion  following from this action are 
\begin{equation}\label{EOMADS} 
 (-z^2\partial_i\partial_i + m^2)\phi + (d-1) z\partial_z\phi - z^2 \partial_z^2\phi=0\;, 
\end{equation} 
where $\partial_i\partial_i$ is the flat space (or boundary) Laplacian, i.e., with indices contracted with the flat metric. 
In order to inspect some simple solutions we make the ansatz $\phi=z^{\Delta}\phi_0$, where $\phi_0$ and $\Delta$ are constant (the latter being referred to as the conformal dimension).  
Insertion into (\ref{EOMADS}) then yields 
a quadratic equation for $\Delta$ whose  solutions are $\Delta_{\pm} = \frac{d}{2} \pm \sqrt{\frac{d^2}{4} + m^2}$, 
where $\Delta_+$ is positive and $\Delta_-$ is negative for $m>0$. 
For more general solutions of (\ref{EOMADS})  that depend on $x$ we impose the boundary condition that they asymptote for $z\rightarrow 0$ 
the solution with 
$\Delta_-$, i.e., 
\begin{equation}\label{asymptoticbehaviorofphi} 
\phi(x,z) = z^{\Delta_-}(\phi_{0}(x) + {\cal O}(z))\; . 
\end{equation}
In other words, we parametrize solutions using $\phi_0(x)$.  Note that we cannot just evaluate these solutions at $z = 0$  since, 
owing to $\Delta_-$ being negative, they become singular. 

In order to now describe the free scalar field on AdS as a chain complex, we define an $L_{\infty}$ algebra with a degree $0$ space of fields $X^0$. We assume that these fields are of the form 
\begin{equation}\label{AdSdecomp}
\begin{split}
\phi(x,z) &= \phi^-(x) z^{\Delta_-}(1 + a_1 z + \cdots ) + \phi^+(x) z^{\Delta_+}(1+b_1 z+\cdots ) \\
&\equiv \phi^-(x,z) + \phi^+(x,z) \, 
\end{split}.
\end{equation}
This is the space of off-shell fields,  
for which  we can define the following projection  to the boundary:
\begin{equation}
p(\phi)(x) := \lim_{z \rightarrow 0} \,z^{-\Delta_-} \phi(x,z) \, .
\end{equation}

To make the action finite, one has to apply \emph{holographic renormalization}, see  e.g.~\cite{Skenderis:2002wp}.  
 In this work, we will ignore this issue, since here  it is only  important that the boundary term obtained when  
integrating by parts in the kinetic term is given by 
\begin{equation}\label{AdSCFTBoundaryTerm}
\begin{split}
\lim_{z \rightarrow 0} \int_{\mathbb{R}^d}  \text d^d x \,  z^{-d + 1} \phi(x,z) \partial_z \phi(x,z) &= \lim_{z \rightarrow 0} \int_{\mathbb{R}^d}  \text d^d x \, z^{-\Delta_-} \phi(x,z) z^{- \Delta_+} z \partial_z \phi(x,z) \\
&= \int_{\mathbb{R}^d} \text d^d x \, \Delta_+ \phi^-(x) \phi^+(x) \, . 
\end{split}
\end{equation}
The first line makes sense for finite values of $z$. In the limit $z \rightarrow 0$, we then make the substitution 
\begin{equation}
z^{-\Delta_+}z \partial_z\phi(x,z) \mapsto z^{-\Delta_+}z \partial_z \phi_+(x,z) \, ,
\end{equation}
which can be justified using holographic renormalization.

The last line in \eqref{AdSCFTBoundaryTerm} suggests 
 that we define $\pi(x) = \Delta_+ \phi^+(x)$ to be the canonical momentum on the boundary. Since $\partial_n = z \partial_z$ is a vector field normal to the boundary and satisfying $g(\partial_n,\partial_n) = 1$, this is very similar to what happens in the scalar field case on ordinary manifolds. Explicitly, the canonical momentum is  obtained as the limit 
 \be
   \pi(x)  : = \lim_{z \rightarrow 0}z^{-\Delta_+}\partial_n \phi^+(x,z)\;. 
 \ee

The space of equations of motion $X^1$ then consists of fields $(\phi^*,\pi^*)$, where $\phi^*$ is a bulk field and $\pi^*$ is a boundary field. We then have a chain complex
\begin{equation}
\begin{tikzcd} 0 \arrow[r] & X^0 \arrow[r,"b_1"] & X^1 \arrow[r] & 0 \, ,
\end{tikzcd}
\end{equation}
where 
 \begin{equation}
  b_1(\phi)  := ((- \Delta + m^2)\phi\,, \; \Delta_+\phi^+)\,. 
 \end{equation} 
We define the cyclic structure as
\begin{equation}
\omega((\phi^*,\pi^*),\phi) := \int_{AdS} \text d^d x \text d z \sqrt{g} \, \phi^*(x,z)\phi(x,z) + \int_{\mathbb{R}^d} \text d^d x \, \phi^-(x) \pi^*(x) \, .
\end{equation}
This correctly reproduces the free action, since
\begin{equation}
\begin{split}
\frac{1}{2} \omega(b_1 \phi,\phi) &= \int_{AdS} \text d^d x \text d z \, \sqrt{g}\phi(-\Delta + m^2)\phi + \int_{\mathbb{R}^d} \text d^d x \,  \Delta_+ \phi^-(x) \phi^+(x) \\
&= \int_{AdS} \text d^d x \text d z \, \sqrt{g} (g^{\mu\nu}\partial_\mu\phi(x)\partial_\nu \phi(x) + m^2 \phi(x)) \, .
\end{split}
\end{equation}

\subsubsection*{Green's Function}

The Green's function $G$ of the kinetic operator $L = -\Delta + m^2$ is given by a certain Gaussian hypergeometric function \cite{Burges:1985qq}.  
The details are not important for us. In the following, we only need its behavior close to the boundary. Explicitly, it satisfies 
\begin{equation}
G(x,z_1; y,z_2) \stackrel{z_1 \rightarrow 0}{\longrightarrow}  c \frac{(z_1z_2)^{\Delta_+}}{(z_2^2 + (x-y)^2)^{\Delta_+}} \, ,
\end{equation}
where $c$ is some normalization constant that is irrelevant for our purposes. 
The Green's function $G$ satisfies Dirichlet boundary conditions in the sense that its weighted boundary value is zero, i.e.
\begin{equation}
p h(\phi^*) := \lim_{z_1 \rightarrow 0} z_1^{-\Delta_-}\int \text d z_2 \text d^d w \, G(x,z_1,y,w) \phi^*(y,w) = 0
\end{equation}
for all fields $\phi^*$. 

The associated inclusion map $i: X^0 \rightarrow X^1$ can be found as in the case of a scalar field on an ordinary manifold. Explicitly, we first compute the boundary symplectic structure via
\begin{equation}
\omega(L\phi_1,\phi_2) - \omega(\phi_1,L\phi_2) = \int_{\mathbb{R}^d} \text d^d x \, \Delta_+(\phi_1^-(x)\phi_2^+(x) - \phi_2^-(x)\phi_1^+(x)) =: \omega_\partial(\phi_1,\phi_2) \, .
\end{equation}
Now recall that, given a boundary field $\phi_0$, the solution $i(\phi_0)$ can be found via
\begin{equation}
\omega(i(\phi_0),\phi^*) = \omega_\partial(\phi_0,\pi h(\phi^*))\, ,
\end{equation}
where 
 \begin{equation} 
   \pi(\phi) = \lim_{\varepsilon \rightarrow 0}(\varepsilon^{-\Delta_-}\phi(x,\varepsilon),\varepsilon^{-\Delta_+}                                                     \varepsilon\partial_\varepsilon\phi(x,\varepsilon)) = (\phi^-(x),\Delta_+\phi^+(x))
 \end{equation}     
    is the projection onto phase space. We find
\begin{equation}
\omega_\partial(\phi_0,\pi h(\phi^*)) = \int_{\mathbb{R}^d} \text d^d x \, \phi_0(x) \int_{AdS} \text d^d y \, \text d z \, \lim_{\varepsilon \rightarrow 0} \varepsilon^{-\Delta_+}\varepsilon\partial_\varepsilon G(x,\varepsilon,y,z) \phi^*(y,z) \, .
\end{equation}
Therefore,
\begin{equation}\label{AdSi0}
\begin{split}
i_0(\phi_0)(x,z) &= \int_{R^d} \text d^d y \, \lim_{\varepsilon \rightarrow 0}  \varepsilon^{-\Delta_+} \varepsilon \partial_\varepsilon G(x,z,y,\varepsilon) \phi_0(y) \\
 &= c\Delta_+ \int_{R^d} \text d^d y \, \frac{z^{\Delta_+}}{(z^2 + (x-y)^2)^{\Delta_+}} \phi_0(y) \, .
\end{split}  
\end{equation}
This indeed satisfies $p_0i_0(\phi_0) = \phi_0$ \cite{Witten:1998qj}. Further, by construction it also satisfies $hL = 1 - i_0p_0$. Therefore, we have a deformation retract
\begin{equation}
\begin{tikzcd}
0 \arrow[r] & X^0 \arrow[d,"p"] \arrow[rr,"-\Delta + m^2"]  & & X^1 \arrow[d] \arrow[r] & 0 \ , \\
0 \arrow[r] & \bar{X}^0 
\arrow[rr] & & 0 \arrow[r] & 0 \ .
\end{tikzcd}
\end{equation}
The space $\bar{X}^0$ consists of fields defined on the conformal boundary of AdS.

\subsubsection*{Cyclic Extension of the Homotopy}

From our experience with the scalar field on a manifold with boundary we know that the projection map in degree one is given by
\begin{equation}
p_1(\phi^*,\pi^*)(x) = \pi^*(x) - \lim_{\varepsilon \rightarrow 0} \varepsilon^{-\Delta_+}\varepsilon\partial_\varepsilon h(\phi^*)(x,\varepsilon) \, .
\end{equation}
The second term on the right hand side computes the canonical momentum of a solution with source $\phi^*$. 
It projects to $\bar{X}^1$, which consists only of the boundary anti-fields $\pi^*$. The inclusion in degree one is the standard one, 
i.e.~$i(\pi^*) = (0,\pi^*)$. From $b_1(\phi) = (L\phi,\Delta_+ \phi_+)$ it follows that
\begin{equation}\label{p1b1ads}
\begin{split}
p_1 b_1(\phi)(x) &= \Delta_+ \phi_+(x) - \lim_{\varepsilon \rightarrow 0} \varepsilon^{-\Delta_+}\varepsilon\partial_\varepsilon hL(\phi)(x,\varepsilon) \\
 &= \Delta_+ \phi_+(x)- \lim_{\varepsilon \rightarrow 0} \varepsilon^{-\Delta_+}\varepsilon\partial_\varepsilon (1 - i_0 p_0)(\phi)(x,\varepsilon) = \lim_{\varepsilon \rightarrow 0}\varepsilon^{-\Delta_+}\varepsilon\partial_\varepsilon i_0 p_0(\phi)(x,\varepsilon) \, ,
\end{split}
\end{equation}
where we used that $\lim_{\varepsilon \rightarrow 0}\varepsilon^{-\Delta_+}\varepsilon\partial_\varepsilon \phi(x,\varepsilon) = \Delta_+ \phi_+(x)$. The differential $c_1$ on the boundary fields should compute the canonical momentum of a solution with boundary value $\phi_0$. Therefore, we define
\begin{equation}
c_1(\phi_0)(x) = \lim_{\varepsilon \rightarrow 0}\varepsilon^{-\Delta_+}\varepsilon\partial_\varepsilon i_0(\phi_0)(x,\varepsilon) \, .
\end{equation}
Using \eqref{p1b1ads}, we then find that $p_1 b_1 = c_1 p_0$, so $p_0$ and $p_1$ define a chain map.
In order to verify that $i_0$ and $i_1$ also define a chain map we compute
\begin{equation}
b_1 i_0(\phi_0) = (0,\lim_{\varepsilon \rightarrow 0} \varepsilon^{-\Delta_+} \varepsilon \partial_\varepsilon i_0(\phi_0)(x,\varepsilon)) = i_1 c_1(\phi_0)\, .
\end{equation}

Finally, we want to extend the homotopy. We extend $h$ so that it is independent of $\pi^*$, i.e. $h(\phi^*,\pi^*) = h(\phi^*)$. We then find
\begin{equation}
\begin{split}
b_1 h(\phi^*,\pi^*) &= (\phi^*,\lim_{\varepsilon \rightarrow 0} \varepsilon^{-\Delta_+}\varepsilon \partial_\varepsilon h(\phi^*)(x,\varepsilon)) = (1 - i_1p_1)(\phi^*,\pi^*) \, .
\end{split}
\end{equation}
As before, we also have $hb_1 = 1 - i_0 p_0$. Therefore, we have in fact a strong deformation retract, given by
\begin{equation}
\begin{tikzcd}
0 \arrow[r] & X^0 \arrow[d,"p_0"] \arrow[r,"b_1"]  & X^1 \arrow[d,"p_1"] \arrow[r] & 0 \ , \\
0 \arrow[r] & \bar{X}^0 \arrow[r,"c_1"] & \bar X^1 \arrow[r] & 0 \ .
\end{tikzcd}
\end{equation}

\subsection{Homological Perturbation Lemma }

We can now apply the perturbation lemma by adding a cubic potential. This corresponds to the  2-bracket:
\begin{equation}
b_2(\phi_1,\phi_2) = \lambda \phi_1 \phi_2 \, .
\end{equation}
To get the boundary action by using the inclusion map we apply the perturbation lemma, which  gives us the perturbed inclusion map, which in degree zero is given by
\begin{equation}\label{AdSPLemma}
i'_0 = (1+hb_2)^{-1} i_0 = \sum_{n \ge 0} (-hb_2)^n i_0 \, .
\end{equation}
Note that all maps have to be lifted to maps on the coalgebra. We will compute this map perturbatively in the coupling constant $\lambda$. We denote this expansion as
\begin{equation}\label{inclusionads}
i'_0 = i_{(0)} + i_{(1)} + i_{(2)} + ... \, .
\end{equation}
In our case, $i_{(n)}$ is the $n$th term of the sum in the formula \eqref{AdSPLemma}.

The homotopy $h$ is lifted by first defining it as a map $\tilde h: (X^\bullet)^{\otimes n} \rightarrow (X^\bullet)^{\wedge n}$ as
\begin{equation}
h(a_1 \otimes \cdots \otimes a_n) = \sum_{k=1}^n (-)^{a_1 + ... + a_k} i p(a_1) \wedge \cdots ip(a_{k-1}) \wedge h(a_k) \wedge a_{k+1} \wedge \cdots \wedge a_n \, .
\end{equation}
We can then restrict it to a map acting on $(X^\bullet)^{\wedge n}$ by using the map $s: (X^\bullet)^{\wedge n} \rightarrow (X^\bullet)^{\otimes n}$, given by
\begin{equation}
s(a_1 \wedge \cdots \wedge a_n) = \frac{1}{n!}\sum_{\sigma \in P(n)} \pm a_{\sigma(1)} \otimes \cdots \otimes a_{\sigma(n)}\, , 
\end{equation}
and then setting  $h := \tilde h \circ s$. Here, $P(n)$ denotes the group of permutations.

We now compute the different terms in the expansion \eqref{inclusionads}. We already know from \eqref{AdSi0} that
\begin{equation}
i_{(0)} (\phi_0) (x_1,z_1)= c \Delta_{+} \int \text d^dx_2   \frac{ z^{\Delta_+} \phi_0(x_2)}{(z_1^2+|x_1-x_2|^2)^{ \Delta_+}} \,.
\end{equation}
 To first order in $\lambda$, we find
\begin{equation}
\begin{split}
i_{(1)} (\phi_0, \psi_0)(x,z) = -\lambda c \Delta_{+} \int d^dx_1 dz_1 & d^dy_1 d^dy_2  \sqrt{g}(x_1,z_1)  G(x,z; x_1,z_1) \\ 
& \cdot \frac{z_1^{2\Delta_+} \phi_0(y_1)\psi_0(y_2)}{\left((z_1^2+|x_1-y_1|^2) (z_1^2+|x_1-y_2|^2) \right)^{\Delta_+}} \, .
\end{split}
\end{equation}

For the contribution quadratic in $\lambda$ we do not  give the explicit expressions but  merely state that the perturbation lemma tells us that
\begin{equation}
i_{(2)} = h b_2 (hb_2(i_0(\phi_0), i_0(\psi_0)), i_0(\rho_0)) + \text{cyclic} \, .
\end{equation}
The three terms included in the cyclic sum are precisely analogous to  $s, t$ and $u$ channels.

\subsection{Witten Diagrams and Correlation Functions}
We will now use the previous results for the inclusion map to get correction terms for the action in dependence of boundary data.
The free theory will give us an action of boundary fields, 
from which we can read off the two point function. The correction terms of orders $\lambda$ and $\lambda^2$ will give us 
the first and second correction terms to the action: the three and four-point functions.

The starting point for all calculations is the surface term of the action, obtained 
by integration by parts of the action: 
\begin{equation}\label{InteractingS}
\begin{split}
S =&\; \frac{1}{2} \int_{AdS} \sqrt{g} \text d^dx \text dz \phi(x,z) \big[(- \Delta + m^2) \phi(x,z) + \frac{\lambda}{2} \phi^2(x,z) \big] \\
&\;-\frac{\lambda}{3}\int_{AdS} \sqrt{g} \text d^dx \text dz \phi^3(x,z) + \frac{\Delta_+}{2} \int_{\mathbb{R}^d} \text d^d x \phi^-(x)\phi^+(x) \, .
\end{split}
\end{equation}
The first line vanishes when we consider fields satisfying the full (bulk) equations of motion. The perturbation lemma is such that
\begin{equation}
i'(\phi_0) := \sum_{n \ge 0} \frac{1}{n!}i_{(n)}(\phi_0,...,\phi_0)
\end{equation}
satisfies the full non-linear equations of motion. Therefore, when computing $\bar{S} = S \circ i'(\phi_0)$, we can drop the first line in \eqref{InteractingS}. Using this together with the computation of $i_{(n \le 3)}$ in the previous section, we can compute the boundary action $\bar{S}$ to quartic order. We can further simplify the computation using the fact that
$\phi^- = p(\phi)$ and $ph = 0$. But all the $i_{(n \ge 1)}$ end in a propagator, so we find that $\phi^-(x)$ reduces to $\phi_0$ in all orders. Only the linear part of $i'$ contributes.

\subsubsection*{Boundary Action}

We get the boundary action $\bar S(\phi_0) = S \circ i'(\phi_0)$ to zeroth order in $\lambda$ from \eqref{InteractingS} by noting that $\phi^-(x) = \phi_0(x)$ and
\begin{equation}
\begin{split}
\Delta_+\phi^+(x) &= \lim_{z \rightarrow 0} z^{-\Delta_+} \partial_z 
z \int \text d^d y \frac{c \Delta_{+} z^{\Delta_+ }}{(z^2+|x-y|^2)^{\Delta_+} } \phi_0(y) \\
&= \int \text d^d y \frac{c \Delta^2_{+} }{|x-y|^{\Delta_+} } \phi_0(y) \, .
\end{split}
\end{equation}
Therefore, to this order we find
\begin{equation}
\bar{S}_{(0)}(\phi_0) = \int \text d^d x \text d^d y \,  \frac{c \Delta_+^2}{|x-y|^{2\Delta_+}} \, .
\end{equation}
By the AdS/CFT dictionary, this is the two-point function of an operator of scaling dimension $\Delta_+$ of the CFT on the boundary. It is customary to represent it as the following Witten diagram 
\\
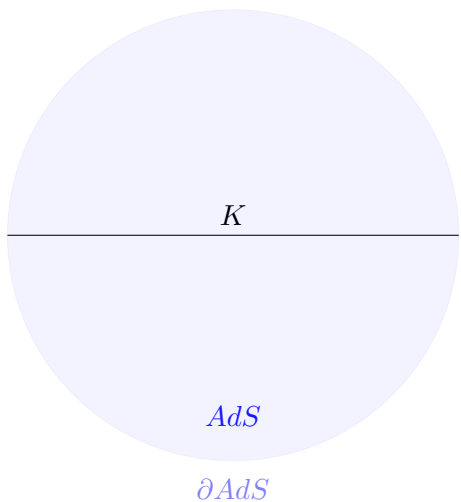
\begin{center}
\begin{tikzpicture}[scale = 1.5]
 \draw (-2,0)   --  (0:2) node[midway,above] {$K$} ;
 \draw[blue,fill=blue, opacity = 0.05] (0,0) circle (2);
\draw [text = blue, opacity = 0.9] (0,-1.6) node {$ AdS$} ;
\draw [text = blue, opacity = 0.5] (0,-2.25) node {$\partial AdS$}  ;
\end{tikzpicture}
\end{center}

Next, we compute $\bar S$ to first order in $\lambda$. As noted before, on-shell the action reduces to
\begin{equation}
\begin{split}
S = -\frac{\lambda}{3}\int_{AdS} \sqrt{g} \text d^dx \text dz \, \phi^3(x,z) + \frac{\Delta_+}{2} \int_{\mathbb{R}^d} \text d^d x \, \phi_0(x)\phi^+(x) \, .
\end{split}
\end{equation}
The first term already is linear in the coupling constant, so only $i_{(0)}$ contributes. On the other hand, in the second term, we need $i_{(1)}$ to compute $\Delta_+ \phi^+$ to first order in $\lambda$. With this, we find
\begin{equation}
\begin{split}
\bar S_{(1)}(\phi_0)=\int \sqrt{g} \text d^dx \text dz \text d^dx_1 &  \text d^dx_2 \text  d^dx_3 \, \phi_0(x_1)\phi_0(x_2)\phi_0(x_3) \\
 & \times \frac{ \lambda c^3 \Delta_{+}^3 z^{3 \Delta_+} }{|(z^2+|x-x_1|^2) |(z^2+|x-x_2|^2)|(z^2+|x-x_3|^2) |^{\Delta_+}}\;. 
\end{split}
\end{equation}
This generates the three-point function of the boundary CFT. The corresponding Witten diagram is
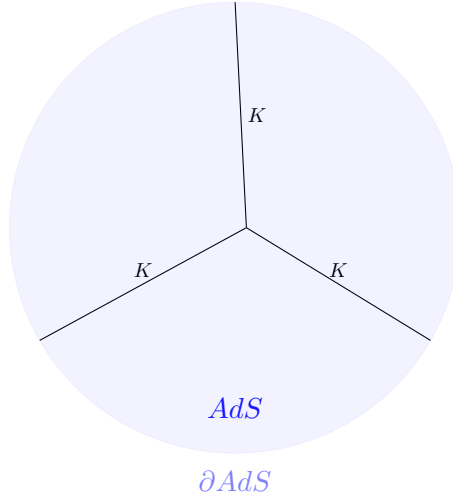
\begin{center}
\begin{tikzpicture}[scale = 1.5]
 \draw (0.1,0) -- (210:2) node[scale=0.7,midway,above] {$K$};
 \draw (0.1,0) -- (-30:2) node[scale=0.7,midway,above] {$K$};
 \draw  (0.1,0) -- (90:2) node[scale=0.7,midway,right] {$K$};
 \draw[blue,fill=blue, opacity = 0.05] (0,0) circle (2);
 \draw [text = blue, opacity = 0.9] (0,-1.6) node {$ AdS$} ;
\draw [text = blue, opacity = 0.5] (0,-2.25) node {$\partial AdS$};
\end{tikzpicture}
\end{center}

Finally, the term quadratic in the coupling constant is given by
\begin{equation}
\begin{split}
\bar S_{(2)}(\phi_0) =   \frac{1}{4!} & \int   \sqrt{g}(z_1)\sqrt{g}(z_2) \text d^d y_1  \text d z_1 \text d^dy_2
\text d z_2 \prod_{i = 1}^4 \text d^d x_i \phi_0(x_i)  \\ 
&\times \frac{3 \lambda^2 c^4 \Delta_{+}^4 z_1^{2 \Delta_+} z_2^{2 \Delta_+} G(y_1,z_1;y_2,z_2) }{|(z_1^2+|y_1-x_1|^2) (z_1^2+|y_1-x_2|^2)(z_2^2+|y_2-x_3|^2)(z_2^2+|y_2-x_4|^2) |^{\Delta_+}}\;, 
\end{split} \, .
\end{equation}
whose  Witten diagram is
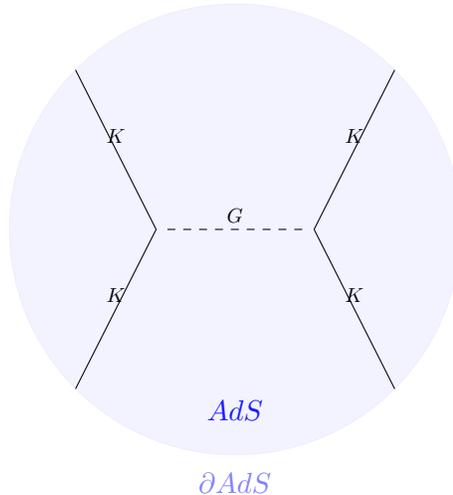
\begin{center}
\begin{tikzpicture}[scale = 1.5]
 \draw (0.7,0) -- (45:2) node[scale=0.7,midway,above] {$K$};
 \draw (0.7,0) -- (-45:2) node[scale=0.7,midway,above] {$K$};
 \draw[dashed] (-0.6,0) -- (0.6,0) node[scale=0.7,midway,above] {$G$};
 \draw (-0.7,0) -- (135:2) node[scale=0.7,midway,above] {$K$};
 \draw (-0.7,0) -- (225:2) node[scale=0.7,midway,above] {$K$};
 \draw[blue,fill=blue, opacity = 0.05] (0,0) circle (2);
  \draw [text = blue, opacity = 0.9] (0,-1.6) node {$ AdS$} ;
\draw [text = blue, opacity = 0.5] (0,-2.25) node {$\partial AdS$}; 
\end{tikzpicture}
\end{center}

\section{Summary and Outlook}

Our goal in this paper was to show that homotopy algebras are an ideal framework to formalize 
holography or the AdS/CFT correspondence, the main reason being  that  the equivalence of structures 
associated to spaces in different dimensions is natural in the world of homotopy algebras. 
More precisely, two spaces of different dimensions may be homotopy equivalent, in which case 
homotopy algebras associated to these spaces are equivalent.

We showed that field theories defined on a `bulk' spacetime with boundary (or conformal boundary in 
the case of AdS) can be encoded in a cyclic $L_{\infty}$ algebra. To this end we refined the dictionary 
between field theories and cyclic $L_{\infty}$ algebras by extending the underlying chain complex 
with proper  boundary components. We then showed, for the examples of scalar field theory and 
Yang-Mills gauge theory, how the passing over to the boundary defines a homotopy retract 
of the underlying chain complex. There is then a homotopy transfer of the  
$L_{\infty}$ algebra encoding  the bulk theory to a `boundary' cyclic $L_{\infty}$ algebra, 
which in turn gives the on-shell action conventionally computed by Witten diagrams. 
Strictly speaking, in order to obtain finite results further boundary terms need to be 
included following the procedure of holographic renormalization \cite{Skenderis:2002wp}.  
Such additional boundary terms may be encoded, for instance, in a further extension of the cyclic $L_{\infty}$ algebra. 
We leave a more detailed 
homotopy algebra formulation of holographic renormalization to future work.

If the AdS/CFT conjecture is correct then the boundary cyclic  $L_{\infty}$ algebra yields the generating functional 
of the correlation functions of the dual field theory defined on the boundary. 
Thus, the results of this paper formalize one arrow of the AdS/CFT dictionary, as depicted in the 
figure of the introduction, thereby providing a promising first step toward a first-principle 
understanding or perhaps even proof of AdS/CFT.

In order to achieve this most ambitious goal, the results of this paper need to be extended in 
various directions. First, the computation of correlation functions starting from a boundary field theory 
needs to be formalized  in terms of homotopy algebras so that one has an $L_{\infty}$ morphism 
from the bulk to the boundary theory (and the other way around).  
Second, the formulation of bulk theories  on AdS in terms of  cyclic $L_{\infty}$ algebras 
needs to be extended to genuine gravity theories, ideally to theories such as type IIB supergravity in 
ten dimensions including all massive Kaluza-Klein modes on, say, $AdS_5\times S^5$. 
As alluded to in the introduction, using techniques from exceptional field theory 
it is clear in principle how to write down these algebras \cite{Hohm:2013pua,Hohm:2013uia,Hohm:2014qga,Baguet:2015sma,Malek:2019eaz}, but it remains to include the boundary components. 
More importantly, as discussed in the introduction, in order to actually prove the AdS/CFT correspondence one 
would presumably need to write down some 
kind of `master' homotopy algebra, say  along the lines of \cite{Bekaert:2012vt},  from which both the boundary and bulk $L_{\infty}$ algebras 
are derivable via some sort of homotopy transfer.

\subsection*{Acknowledgements}

We would like to thank Alex Arvanitakis, Roberto Bonezzi, Tomas Codina, Felipe Diaz-Jaramillo, Owen Gwiliiam, Allison Pinto, Ivo Sachs and Barton Zwiebach for useful discussions and collaborations on related topics.

This work is funded   by the European Research Council (ERC) under the European Union's Horizon 2020 research and innovation programme (grant agreement No 771862).

\appendix

\section{Technical Details  for Yang-Mills Theory with Boundary} 

\subsection{Cyclic Extension in Yang-Mills Theory}

In this appendix we discuss some technical details of the cyclic extensions of the Yang-Mills brackets   given in sec.~\ref{sec:YMproducts}.

We begin by noticing from \eqref{Acycfailure} that
\begin{equation}
\omega(b_1A_1,A_2) + \int_{\partial M} A_2 \wedge \star \text d A_1 = \omega(b_1A_2,A_1)  + \int_{\partial M} A_1 \wedge \star \text d A_2  \; ,
\end{equation}
so the combination on the left-hand side (or equivalently the right-hand side) is symmetric. This is also clear from the fact that \begin{equation}
\frac{1}{2}\omega(b_1A,A) + \frac{1}{2}\int_{\partial M} A \wedge \star \text d A
\end{equation}
is the kinetic term of the classical action $S_0$. Similarly, from \eqref{ccycfailure} we find that
\begin{equation}
\omega(b_1 A^*, c) + \int_{\partial M} c \wedge A^* = -\omega(b_1 c,A^*) \
\end{equation}
is graded symmetric.

In the scalar field case, we only considered interactions without derivatives. On the other hand, the cubic vertex in Yang-Mills theory has a derivative. This derivative also leads to a failure of $b_2$ being cyclic. We compute 
\begin{equation}
\begin{split}
\omega(b_2(A_1,A_2),A_3) &= \int_M A_3 \wedge (\text d \star [A_1,A_2] + [A_2,\star \text d A_1] + [A_1,\star \text d A_2]) \\
						&= \int_M  A_1 \wedge [A_2,\star\text d A_3] - \int_{\partial M}A_3 \wedge \star [A_1,A_2] \\
						&\quad + \int_M \text A_1 \wedge \text d\star[A_2,A_3] + \int_{\partial M} A_1 \wedge \star[A_2,A_3] \\
						& \quad + \int_M A_1 \wedge [A_3,\star \text d A_2] \\
					&= \omega(b_2(A_2,A_3),A_1) - \int_{\partial M}A_3 \wedge \star [A_1,A_2] + \int_{\partial M} \star[A_2,A_3] \wedge A_1 \; .
 \end{split}
\end{equation}
Rewriting this equation, we find that
\begin{equation}
\omega(b_2(A_1,A_2),A_3) +  \int_{\partial M}A_3 \wedge \star [A_1,A_2] = \omega(b_2(A_2,A_3),A_1) + \int_{\partial M} A_1 \wedge \star[A_2,A_3] \; .
\end{equation}
As before, we find that the above defines a symmetric product. Again, this is also obvious from the fact that the left-hand side is equal to the cubic part of the classical action $S_0$.

We now have all the information we need to define an actual cyclic $L_\infty$ algebra. We extend the space of equations of motion by including a space of boundary fields $\Omega^{n-2}(\partial M) \otimes \mathfrak{g}$, whose elements we call $B^*$. Further, we extend the space of Noether identities by a boundary part $\Omega^{n-1}(\partial M) \otimes \mathfrak{g}$, whose fields we call $\eta^*$. We then define the extended complex
\begin{equation}
\begin{tikzcd}
\Omega^0(M)\arrow[r,"b_1'"] & \Omega^1(M) \arrow[r,"b_1'"] & \Omega^{n-1}(M) \oplus \Omega^{n-2}(\partial M)\arrow[r,"b_1'"] & \Omega^n(M)\oplus \Omega^{n-1}(\partial M) \ ,
\end{tikzcd}
\end{equation}
with 
\begin{equation}
b_1'(c) = \text d c \ , \qquad b_1'(A) = (\text d \star \text d A,i^*\star \text d A) \ , \qquad b_1'(A^*,B^*) = (\text d A^*,i^* A^* - \text d B^*) \,, 
\end{equation}
where $i^*$ is the pullback w.r.t.~the inclusion map from the boundary to the bulk. 
We also extend the inner product via
\begin{equation}
\begin{split}
\omega'(c,(c^*,\eta^*)) &= \int_M c \wedge c^* - \int_{\partial M} i^* c \wedge \eta^* \; , \\
\omega'((A^*,B^*),A) &= \int_M A \wedge A^* + \int_{\partial M} i^* A \wedge B^* \ , 
\end{split}
\end{equation}
for which one finds  
\begin{equation}
\begin{split}
\omega'(b_1'A_1,A_2) &= \omega(b_1' A_2,A_1) = \int_M \text d A_1 \wedge \star \text d A_2 \ , \\
\omega'(b_1'c,(A^*,B^*)) &= -\omega'(b_1'(A^*,B^*),c) = -\int_M \text d c \wedge A^* - \int_{\partial M} i^* \text d c \wedge B^* \ .
\end{split}
\end{equation}
Therefore, we correctly reproduce the classical free action. The BV extended action gets the additional term
\begin{equation}
- \int_{\partial M} i^* \text d c \wedge B^* \ .
\end{equation}

For the cubic vertex, we found that the naive product is also not cyclic. We therefore have to extend the product $b_2$ to
\begin{equation}
b_2'(A_1,A_2) = (b_2(A_1,A_2), i^*\star[A_1,A_2]) \in (\Omega^{n-1}(M)\oplus \Omega^{n-2}(\partial M)) \otimes \mathfrak{g} \; .
\end{equation}
One  then finds 
\begin{equation}
\omega'(b_2'(A_1,A_2),A_3) = \int_M \text d A_1 \wedge \star [A_2,A_3] + \text{cyclic} \; ,
\end{equation}
which is manifestly  symmetric. We are not done, however, since 
we still need to make sure that $b_1'$ is a derivation with respect to $b_2'$. We compute 
\begin{equation}
\begin{split}
b_1' b_2'(A_1,A_2) 	&= (b_1 b_2(A_1,A_2),i^* b_2(A_1,A_2) - i^* \text d \star [A_1,A_2]) \\
					&= (-b_2(b_1A_1,A_2) - b_2(A_1,b_1 A_2), [i^*A_1,i^* \star \text d A_2] + [i^* A_2, i^* \star \text d A_1]) \, .
\end{split}
\end{equation}
We see that in the second component of the above expression, $\star \text d A_i$ is the boundary contribution of $b_1'(A_i)$. Therefore, we set
\begin{equation}
b_2'(A,(A^*,B^*)) = (b_2(A,A^*),-[i^* A, B^*]) \ .
\end{equation}
A priori, it is not guaranteed that $b_1'$ is a derivation of that product. However, the output of $b_2'$ in that case already has top degree, and therefore $b_1$ acts trivially. So the derivation property is trivially satisfied.

We extended $b_2'$ so that it also pairs fields with boundary anti-fields $B^*$. For this reason, it is no longer guaranteed that it is cyclic. Therefore, we need to compute
\begin{equation}
\begin{split}
\omega'(b_2'(A,(A^*,B^*)),c) 	&= \omega'(([A,A^*],-[i^*A,B^*]),c) \\
&= - \int_M [A,A^*] \wedge c - \int_{\partial M} [i^*A,B^*] \wedge i^*c \\
								&= \int_M A \wedge [c,A^*] + \int_{\partial M} i^* A \wedge [i^*c,B^*] \\
								&= -\int_M A \wedge [A^*,c] - \int_{\partial M} i^* A \wedge [B^*,i^*c]\\
								&= \omega'(b_2'((A^*,B^*),c),A) \ ,, 
\end{split}
\end{equation}
where the final equation is the statement of cyclicity  for  $b_2'$. 
This is hence satisfied if we  define
\begin{equation}
b_2'((A^*,B^*),c) = (-[A^*,c], -[B^*,i^* c]) \ .
\end{equation}
Once more, we need to check the derivation property of $b_1'$: 
\begin{equation}
\begin{split}
b_1' b_2'((A^*,B^*),c) 	&= (b_1b_2(A^*,c), -i^*[A^*,c]+\text d[B^*,i^* c]) \\
						&= (-b_2(b_1 A^*,c) + b_2(A^*,b_1 c), -[i^* A^* - \text d B^*,i^* c] - [\text i^* d c,B^*]) \\
						&= b_2'((A^*,B^*),b_1' c) -(b_2(b_1 A^*,c),[i^* A^* - \text d B^*,i^* c])\,. 
\end{split}
\end{equation} 
From this we see that we need to define
\begin{equation}
b_2'((c^*,\eta^*),c) = (b_2(c^*,c),[\eta^*,i^*c]) \ .
\end{equation}

We now have everything we want. $b_1'$ and $b_2'$ are cyclic and $b_1'$ is a derivation of $b_2'$. Further, one can show that $b_2'$ satisfies the Jacobi identity up to homotopy $b_3'$, if we set
\begin{equation}
b_3'(A_1,A_2,A_3) = (b_3(A_1,A_2,A_3),0) \, .
\end{equation}

\subsection{Homotopy Retract in Yang-Mills Theory}

\label{App:YMHomotopy}

Here  we give a more in-depth derivation of the homotopy given in section \ref{Sec:YMHomotopy}. On free equations of motion, the homotopy should be the Green's function. We are looking for solutions to 
\begin{equation}\label{YMwithsource}
\text d \star \text d A = A^* \: .
\end{equation}
To solve this, we use Lorenz gauge $\delta A = 0$. In this case, the equation of motion for $A$ can be written as 
\begin{equation}\label{HarmonicYM}
\Delta A = \star A^* \; .
\end{equation}
Note that this  equation does not require $A^*$ to be conserved ($\text d A^* = 0$), 
unlike the original Yang-Mills equation \eqref{YMwithsource}. However, we claim that if we have a solution to \eqref{HarmonicYM} together with $i^*\delta A = 0$,  
then $A$ satisfies Lorenz gauge if and only if $A^*$ is conserved. 

To prove this, suppose we have a solution to \eqref{HarmonicYM} with $i^* \delta A = 0$. We first assume that $A$ satisfies Lorenz gauge. Then, the equation \eqref{HarmonicYM} reduces to the original Yang-Mills equation $\text d \star \text d A = A^*$. From this it immediately follows that $\text d A^* = 0$. On the other hand, assume that $A^*$ is conserved. From \eqref{HarmonicYM}, we then deduce that
\begin{equation}\label{conservationislorenz}
\delta \text d \delta A = 0 \; .
\end{equation}
We define $f = \delta A$. From \eqref{conservationislorenz}, we then find
\begin{equation}
\delta \text d f = \Delta f = 0 \ ,
\end{equation}
where we used the fact that $f$ is a zero form (i.e.~a Lie algebra valued function). Now $i^* f = i^*\delta A = 0$ by assumption. But this means that $f$ should be a harmonic functions vanishing on the boundary.  But from this it follows that $f = 0$ everywhere and therefore $0 = \delta A$.

We now turn to existence of solutions, where we will solve \eqref{HarmonicYM} instead of \eqref{YMwithsource}. Lemma 3.4.7 in \cite{Schwarz1995HodgeD} states that the equation $\Delta A = \star A^*$, with boundary condition  $i^*(A) = a_0, i^*\delta A = 0$, has a solution, if and only if 
$A^* \in (\mathcal{H}_N^{d-1}(M))^{\perp}$, where $\perp$ is the orthogonal with respect to the inner product $(-,-)$ defined in \eqref{YMinnerproduct}. The solution is unique up to $A \in \mathcal{H}^1_D(M)$. In general, we can project $A^*$ to $(\mathcal{H}_N^{d-1}(M))^{\perp}$ first using the Hodge decomposition
\begin{equation}
\Omega^{n-1}(M) = \text d \Omega^{n-2}_D(M) \oplus \delta \Omega^{n}(M) \oplus \mathcal{H}_N^{d-1}(M) \oplus \mathcal{H}_{ex}^{d-1}(M) \; ,
\end{equation} 
and then solving  the problem such that $A\in (\mathcal{H}^1_D(M))^{\perp}$. Taking the solution with $i^*A = 0$, this defines a Green's function $G: \Omega^{n-1}(M) \rightarrow \Omega^{1}(M)$, such that
\begin{equation}
\Delta G_1(A^*) = \star(1-I_1 P_1) A^* \ ,
\end{equation}
where $P_1: \Omega^{n-1}(M) \rightarrow \mathcal{H}_N^{n-1}(M)$ is the projector to $\mathcal{H}_N^{n-1}(M)$ using the Hodge decomposition and $I_1$ is the inclusion of that space. Just like in the scalar field case, we take the Green's function satisfying Dirichlet boundary conditions. Note that here $A$ being Dirichlet means that $i^* A = 0$, and not $A|_{\partial M} = 0$. There is also a solution to the latter boundary condition under different assumptions. But we do not want to solve for the normal component, since it will be determined by the `constraint equation'.\footnote{This is a constraint equation if we think of the direction normal to the boundary as ``time''.}

Apart from $G_1$, we will also use the scalar Green's function $G_0$, which we encountered in the section on scalar field theory. With this, we now have all the tools to define the homotopy. First of all, on $V^1 = \Omega^{n-1}(M) \otimes \mathfrak{g}$, we define
\begin{equation}
h(A^*) = G_1(1-I_2 P_2)(A^*) \ ,
\end{equation} 
where $P_2$ and $I_2$ are the projection and inclusion onto and from $\delta\Omega^{n}_N(M)$ coming from the Hodge decomposition. We use them, since as we saw before, this ensures that $h(A^*)$ satisfies Lorenz gauge. We further define
\begin{equation}
h(A) = G_0\delta A \ 
\end{equation}
to  find
\begin{equation}\label{gaugehomotopy}
h b_1 c = G_0 \delta \text d c = G_0 \Delta c = 1-i_{-1} p_{-1} \; ,
\end{equation}
where $p_{-1} = i^*$ and $i_{-1}$ associates a solution $\Delta c = 0$ to a given boundary value $c_0$. The relation \eqref{gaugehomotopy} is exactly as in the scalar field case. 
 So the homotopy interpolates between a bulk gauge parameter to a boundary gauge parameter (the word \emph{gauge} will be justified below). On the other hand, to see what the homotopy does on gauge fields, we compute $hb_1(A)$ and $b_1h(A)$ separately. First of all, we have $b_1h(A) = \text d G_0 \delta A$. We claim that this projects $A$ to $\text d a \in \text d \Omega_D(M)$, i.e.~to a `pure gauge' component. To see this, we use the Hodge decomposition \eqref{HodgeDecomp} to find
\begin{equation}
b_1h(A) = \text d G_0 \delta \text d a \; .
\end{equation}
From the scalar field case, we know that $G_0 \delta \text d = 1 - i_{-i}p_{-1}$. But from $a \in \Omega^0_D(M)$, it follows that $p_{-1}(a) = 0$. So we find $G_0 \delta \text d a = a$ and therefore $b_1h(A) = \text d a$. On the other hand,
\begin{equation}
hb_1(A) = G_1(1-I_2P_2)\text d \star \text d A = G_1 \text d \star \text d A \; ,
\end{equation}
where we used that $P_2 \text d = 0$, which follows from the fact that the image of $\text d$ is orthogonal to $\delta \Omega_N(M)$. We now show that $b_1(1-hb_1)(A) = 0$. To this end one computes 
\begin{equation}
\begin{split}
b_1 h b_1 (A) 	&= \text d \star \text d G_1\text d \star \text d A = -(-)^n\star\Delta G_1 \text d \star \text d A = -(-)^n (1-I_1 P_1) \star \delta \text d A \\
				&= \text d \delta \text d A = b_1(A) \ .
\end{split}
\end{equation}
Here we used that $P_1 \delta = 0$, which is true since the image of $\delta$ is orthogonal to $\mathcal{H}_N^{n-1}$. From the above it follows that $b_1(1-hb_1) = 0$, so $hb_1$ projects to solutions to the equations of motion. 

Recall that our original goal is to find a homotopy to a boundary theory. This will not be entirely possible, since there is a topological obstruction in $(\mathcal{H}^1_D)^\perp$, which is part of the space of solutions not removeable by a choice of gauge. For this reason, we first define $P_3: \Omega^1(M) \rightarrow \mathcal{H}^1_D(M)$ using the Hodge decomposition and $I_3: \mathcal{H}^1_D(M)\rightarrow \Omega^1(M)$,  the natural inclusion. We project the gauge fields $A$ via $p_0(A) = (i^* A,P_3(A))$. To define our inclusion, we associate to a boundary configuration $a_0 \in \Omega^1(\partial M)$ the unique solution to $\Delta A = 0$, such that $i^* A = a_0, i^* \delta A = 0, A \in (\mathcal{H}_D^1)^\perp(M)$.
Since all boundary conditions 
are linear, the map $r: \Omega^1(\partial M) \rightarrow \Omega^1(M)$ associating to a boundary one-form $a_0$ a harmonic bulk form is linear. We can then define the linear map $i_0: \Omega^1(\partial M) \oplus \mathcal{H}^1_D(M) \rightarrow \Omega^1(M)$, where $i_0 = r + I_3$. We recall also that $i^* \delta A = 0$ and $\delta\Delta A = 0$ implies $\delta A = 0$, so $A$ satisfies Lorenz gauge. Therefore $\Delta A = \star \text d \delta \text d A = 0$, so $A$ solves the linear equations of motion of Yang-Mills theory.  By construction, we have $p_0 i_0 = 1$. To show the homotopy property
\begin{equation}\label{YMHomotopy}
(1- i_0 p_0)(A) = (b_1 h + h b_1)(A) \ ,
\end{equation}
we prove that $(1-b_1 h - h b_1)(A)$ is the sum of the topological part of $A$, i.e.~$I_3P_3(A) \in \mathcal{H}_D^1(M)$, plus a solution $\mathcal A$ with the boundary conditions
\begin{equation}\label{1formbc}
i^* \mathcal{A} = \mathcal{A} \,, \qquad i^*\delta \mathcal A = 0 \ , \qquad \mathcal{A} \in (\mathcal{H}_D^1(M))^\perp \ .
\end{equation} 
Since $i_0p_0(A)$ has the same property, it follow from uniqueness of solutions that \eqref{YMHomotopy} holds.

We begin by noting that since $b_1 h$ projects to $\Omega^1_D(M)$, we have that $\delta b_1 h = \delta$. Further, the image of $h$ satisfies Lorenz gauge, so $\delta h = 0$. Therefore, $\delta(1-b_1 h - h b_1)(A) = \delta A - \delta A = 0$. Further, $i^* b_1 h (A) =i^* \text d h (A) = \text d i^* h(A) = 0$, where we used that $i^*$ commutes with $\text d$ and that $h(A)$ is a function satisfying Dirichlet boundary conditions. Also, by construction we have that $i^*hb_1(A) = 0$, since we chose $G_1$ such that $i^* G_1 = 0$. Hence, $i^*(1-hb_1 - b_1 h)(A) = i^* A$. Also, by construction we have $P_3h b_1(A) = P_3 b_1 h(A) = 0$. It follows that $I_3P_3(1-hb_1 - b_1 h)(A) = I_3P_3(A)$ is the part of $A$ in $\mathcal{H}_D^1(M)$. Further, $(1- I_3P_3)(1-hb_1 - b_1 h)(A)$ is a solution with boundary conditions \eqref{1formbc}. But this proves that
\begin{equation}
(1 - hb_1 - b_1)(A) = I_3P_3(A) + (1-I_3P_3)(1 - hb_1 - b_1)(A)
\end{equation}
is exactly the split we are looking for. This finishes the proof of \eqref{YMHomotopy}.

Before going any further, let us see summarize what we already got. We are half-way in the construction of a homotopy to a boundary theory (plus a topological part). We have a homotopy equivalence
\begin{equation}
\begin{tikzcd}
0 \arrow[r] & \Omega^0(M) \arrow[r,"\text d"] \arrow[d,"p_{-1}"] & \Omega^1(M) \arrow[r,"\text d \star \text d"] \arrow[d,"p_0"] & \Omega^{n-1}(M) \arrow[r,"\text d"] \arrow[d,"?"] & \Omega^n(M) \arrow[r] \arrow[d,"?"] & 0 \\
 0 \arrow[r] & \Omega^0(\partial M) \arrow[r,"\text (0{,}d)"] &  \Omega^1(\partial M) \oplus \mathcal{H}_D^1(M) \arrow[r,"?"] & ? \arrow[r,"?"] & ? \arrow[r] & 0
\end{tikzcd}
\end{equation}
All question marks still need to be determined. However, from the first two non-trivial terms in the complex we already see that the bulk theory on $M$ constracts to a boundary theory on $\partial M$ plus the topological term $\mathcal{H}_D(M)$. The fact that $p_{-1}$ and $p_0$ combine into a chain map commuting with $\text d$ immediately follows from $i^*$ commuting with $\text d$.

We continue by constructing the last piece of the homotopy. This should be a map $h_2: \Omega^n(M) \rightarrow \Omega^{n-1}$ and further we want it to be symmetric with respect to $\omega$. In the end, only the extension of $h$ to the cyclic theory has to be symmetric in order to get a cyclic homotopy transfer needs to be symmetric. However, an $h$ symmetric with respect to $\omega$ has a symmetric extension. Symmetry with respect to $\omega$ means that
\begin{equation}\label{SymmetricH}
\omega(h_2 c^*,A) = \omega(c^*,h_0 A) \ .
\end{equation}
From the fact that $h_0(A) = G_0 \delta$ and $G_0$ are symmetric with respect to $\omega(\phi_1,\star \phi_2)$ (which we know  from our study of the scalar field), we find from \eqref{SymmetricH} that $h_2 = (-)^n \delta \star G_0 \star$. Note that $G_n = (-)^n\star G_0 \star$ defines a Green's function on top forms. Further, since $\star$ maps forms satisfying Dirichlet boundary conditions to forms satisfying Neumann boundary conditions, it follows that $G_n$ satisfies Neumann boundary conditions, which for top forms means that they just vanish at the boundary.

We first compute the effect of $h_2$ on $c^* \in \Omega^n(M)$. We find
\begin{equation}
b_1 h_2 (c^*) = (-)^n\text d \delta \star G_0 \star c^* = \star \delta \text d G_0 \star c^* = \star^2 c^* = c^* \ . 
\end{equation}
Hence $h_2$ contracts $\Omega^n(M)$ to zero. This is consistent with the fact that $\Omega^n(M)$ has no cohomology on a compact and connected manifold with boundary. On the other hand, for an $A^* \in \Omega^{n-1}(M)$, we find 
\begin{equation}
b_1 h_1(A^*) = (1-I_1P_1 - I_2P_2)(A^*) \ ,
\end{equation}
where $I_1P_1 + I_2 P_2$ projects $A^*$ to $\delta \Omega_N^{n-1}(M) \oplus \mathcal H^{n-1}_N(M)$. We also want to show that $h_2 b_1 = I_2P_2$. For this it is sufficient to show that $h_2 b_1$ is zero on $\text d \Omega_D^{n-2} \oplus \mathcal{H}^{n-1}(M)$, which follows from the fact that $b_1$ acts trivially on these, and further that it acts on $\delta \Omega_N^{n}$ as the identity. So suppose $\delta c^* \in \delta\Omega^{n}_N(M)$. We then have
\begin{equation}
h_2 b_1 (\delta c^*) =  (-)^n\delta \star G_0 \star \text d \delta c^* = \delta \star G_0  \delta \text d \star  c^* = \delta \star^2 c^* = \delta c^* \ .
\end{equation}
Here, we used that $\star c^*$ satisfies Dirichlet boundary conditions, which follows from $c^*$ satisfying Neumann boundary conditions (recall that $\mathbf{t}\star = \star\mathbf{n}$), and that $G_0 \delta \text d = 1$ on scalar fields satisfying Dirichlet boundary conditions. We therefore conclude that $h_2 b_1 = I_2 P_2$. Altogether, we find
\begin{equation}
(b_1 h_1 + h_2 b_1)(A^*) = A^* - I_1 P_1(A^*) =: 1 - i_1 p_1(A^*) \ .
\end{equation}
Note  that, acting on $A^*$, $I_1, P_1$ define the complete inclusion and projection, 
and so 
we switched to the lower case alphabet in agreement with our previous notation. 
The above shows that the homotopy contracts the space of equations $\Omega^{n-1}(M)$ to $\mathcal{H}_N^{n-1}(M)$, which is purely topological.

We now are now able to fill the question marks in the commutative diagram describing the homotopy. Altogether, we find
\begin{equation}
\begin{tikzcd}
0 \arrow[r] & \Omega^0(M) \arrow[r,"\text d"] \arrow[d,"p_{-1}"] & \Omega^1(M) \arrow[r,"\text d \star \text d"] \arrow[d,"p_0"] & \Omega^{n-1}(M) \arrow[r,"\text d"] \arrow[d,"p_1"] & \Omega^n(M) \arrow[r] \arrow[d,"0"] & 0 \\
 0 \arrow[r] & \Omega^0(\partial M) \arrow[r,"\text (0{,}\text d)"] & \mathcal{H}_D^1(M) \oplus \Omega^1(\partial M) \arrow[r,"0"] &  \mathcal{H}_N^{n-1}(M) \arrow[r,"0"] & 0 \arrow[r] & 0
\end{tikzcd}
\end{equation}
In the first two non-trivial terms, we find a boundary theory consisting of gauge parameters and gauge fields. On top of that, in degree zero and one there are topological terms $\mathcal H^1_D(M)$ and $\mathcal{H}^{d-1}_N(M)$. Note that these two are Hodge dual to each other, so if one is zero, so is the other.

\end{document}